\documentclass[journal,twoside,web]{ieeecolor}
\usepackage{generic}
\usepackage[noadjust]{cite}
\usepackage{amsmath,amssymb,amsfonts,nccmath}
\usepackage{graphicx}
\graphicspath{{./figures/}}
\usepackage{textcomp}
\def\BibTeX{{\rm B\kern-.05em{\sc i\kern-.025em b}\kern-.08em
    T\kern-.1667em\lower.7ex\hbox{E}\kern-.125emX}}
\usepackage{latexsym}
\usepackage[utf8]{inputenc} %
\usepackage[T1]{fontenc}    %
\usepackage{booktabs}       %
\usepackage{nicefrac}       %
\usepackage{microtype}      %
\usepackage{caption}
\usepackage{subcaption}
\usepackage{xr}
\usepackage{bbold}
\usepackage{siunitx}
\DeclareSIUnit\RPM{rpm}
\usepackage{mathtools}
\usepackage{isomath}
\usepackage{float}
\usepackage{bm}
\usepackage{color}
\usepackage{nicefrac}
\usepackage{csquotes}
\usepackage{tabularx}   %
\newcommand{\bigcell}[2]{\begin{tabular}{@{}#1@{}}#2\end{tabular}}

\begin{document}
\bstctlcite{IEEEexample:BSTcontrol}
\title{Global Sensitivity Analysis of Four Chamber Heart Hemodynamics Using Surrogate Models}
\author{Elias Karabelas, Stefano Longobardi, Jana Fuchsberger, Orod Razeghi, Cristobal Rodero, Marina Strocchi, Ronak Rajani, Gundolf Haase, Gernot Plank and Steven Niederer
\thanks{
S. Niederer acknowledges support from the UK Engineering and Physical Sciences Research Council (grant nos. EP/M012492/1, NS/A000049/1 and EP/P01268X/1), the British Heart Foundation (grant nos. PG/15/91/31812, PG/13/37/30280, SP/18/6/33805), US National Institutes of Health (grant no. NIH R01-HL152256), European Research Council (grant no. ERC PREDICT-HF 864055), Wellcome Trust (grant no. WT 203148/Z/16/Z) and Kings Health Partners London National Institute for Health Research (NIHR) Biomedical Research Centre, and UK HPC resources ARCHER.}
\thanks{G. Haase and G. Plank acknowledge support from grants nos. F3210-N18 and I2760-B30 from the Austrian Science Fund (FWF) and by BioTechMed-Graz (Grant No. Flagship Project: ILearnHeart).}
\thanks{C. Rodero acknowledges support from from the European Union’s Horizon 2020 Research and innovation programme under the Marie Sklodowska-Curie Grant Agreement No 764738.}
\thanks{We further acknowledge support by NAWI Graz and by the PRACE project \enquote{71138: Image-based Learning in Predictive Personalized Models of Total Heart Function} for awarding us access to the Austrian HPC resources VSC4.}
\thanks{
E. Karabelas, J. Fuchsberger, and G. Haase are with the Department of Mathematics and Scientific Computing, University of Graz, Graz, AT}
\thanks{
S. Longobardi, C. Rodero, M.Strocchi, and S. Niederer (corresponding author) are with the Cardiac Electromechanics Research Group, School of Biomedical Engineering and Imaging Sciences, King’s College London, London, UK (e-mail: steven.niederer@kcl.ac.uk)}
\thanks{
O. Razeghi is with the Research IT Services Department, University College London, London, UK}
\thanks{R. Rajani is with the Department of Adult Echocardiography, Guy’s and St Thomas’ Hospitals NHS Foundation Trust, London, UK}
\thanks{G. Plank is with the Gottfried Schatz Research Center (for Cell Signaling, Metabolism and Aging), Division Biophysics, Medical University of Graz, Graz, AT}
\thanks{E.~Karabelas and S.~Longobardi contributed equally to this paper.}
\thanks{\textbf{This work has been submitted to the IEEE for possible publication. Copyright may be transferred without notice, after which this version may no longer be accessible.}}
}

\maketitle

\begin{abstract}
Computational Fluid Dynamics (CFD) is used to assist in designing artificial valves and planning procedures, focusing on local flow features.
However, assessing the impact on overall cardiovascular function or 
predicting longer-term outcomes may requires more comprehensive whole heart CFD models. 
Fitting such models to patient data requires numerous computationally expensive simulations,
and depends on specific clinical measurements to constrain model parameters, 
hampering clinical adoption. 
Surrogate models can help to accelerate the fitting process while accounting for the added uncertainty.
We create a validated patient-specific four-chamber heart CFD model based on the Navier-Stokes-Brinkman (NSB) equations 
and test Gaussian Process Emulators (GPEs) as a surrogate model 
for performing a variance-based global sensitivity analysis (GSA). 
GSA identified preload as the dominant driver of flow
in both the right and left side of the heart, respectively. 
Left-right differences were seen in terms of vascular outflow resistances,
with pulmonary artery resistance having a much larger impact on flow
than aortic resistance. %
Our results suggest that GPEs can be used to identify parameters 
in personalized whole heart CFD models, 
and highlight the importance of accurate preload measurements.
\end{abstract}


\section{Introduction}
\label{sec:introduction}
\IEEEPARstart{V}{alvular} heart disease is a growing problem with limited pharmacological therapies \cite{Coffey2014}. 
Patients with valvular malfunctions are at high risk of developing cardiovascular diseases (CVD) \cite{owidcausesofdeath}.
Valve treatments rely on invasive surgery or catheter-based implanted valves \cite{Manolis2017}. 
Choosing the best option for each patient remains a challenge \cite{Baumgartner2017}.

However, our understanding of how valvular diseases affect the heart and cardiovascular system as a whole remains incomplete. 
Mechanistic models \cite{Niederer2018} encapsulate our knowledge of physiology and the underlying fundamental laws of physics. 
They provide a framework to integrate experimental and clinical data, 
enabling the identification of mechanisms and/or the prediction of outcomes, even under unseen scenarios without the need for retraining \cite{Davies2016}.
Computational fluid dynamics (CFD) is routinely used for designing valves \cite{Mittal2016} and guiding implantation planning \cite{Brouwer2018}. 
These simulations focus on modeling local blood flow across the valve 
and do not consider blood flow in the wider heart. 
Simulating blood flow in the whole heart can be important when estimating pressure gradients 
in the left ventricular outflow tract in transcatheter mitral valve implants (TMVI) \cite{Blanke2017}, 
or when considering ventricle size in transcatheter aortic valve implants (TAVI) \cite{Saito2021}.
However, patient-specfic simulations of blood flow in the whole heart requires parameters and boundary conditions to be tuned to an individual, requiring numerous expensive simulation.
There is a need to reduce the computational cost of simulations and to focus simulations on tuning important parameters.
Previous studies have performed local sensitivity analysis in simplified models, see for example \cite{Ellwein2007,GarcaIsla2018}, however, these fail to provide an estimate of global and multi-factorial sensitivity. 
Identifying the key parameters that need to be personalized will both focus clinical measurements of key patient phenotypes and reduce the parameter space that needs to be explored to personalize the models. 

The gold standard for modeling valves casts blood-valve interaction 
as a transient fluid–structure interaction (FSI) problem.
Recent advances \cite{Astorino2009,Wenk2010,Terahara2020} show the potential of fully coupled FSI models.
However, computational costs and  patient-specific parametrization \cite{Stevanella2011} still pose major obstacles, hindering a swift clinical translation.
Immersed boundary methods (IBM) \cite{Peskin1972} 
have proven to be a promising alternative,
combining computational efficiency, ease of implementation, and numerical stability \cite{Mittal2005}, especially when applied to heart valve modeling \cite{Astorino2012,Votta2013,Chnafa2014}.

In this study we create and validate a patient-specific model of blood flow across the four chambers of the heart using the residual-based variational multiscale formulation (RBVMS) \cite{Bazilevs2007} of the arbitrary Lagrangian-Eulerian Navier-Stokes-Brinkman equations (ALE-NSB) \cite{Fuchsberger2021, Daub2020}.
We test the ability of machine learning-based GPEs, 
which approximate the model and estimate the uncertainty in the approximation, to provide a low-cost surrogate for the full physics-based model. 
Using GPEs, we perform a variance-based GSA over parameters 
governing flow in the left and right heart 
to determine which of those are most important and need to be accurately personalized for patient-specific predictions.
\section{Methods}
\label{sec:methods}
\subsection{Ethics Declaration}
This study uses a fully anonymized data set collected at Guy's and St Thomas' Hospital, London, United Kingdom, as part of standard of care.
\subsection{Data Acquisition}\label{sec:data_gen}
The patient received a ECG-gated cardiac CT angiography.
Clinically indicated MDCT was performed as the standard of care using the hospital's 3$^\text{rd}$ generation dual-source CT system (SOMATOM Force, Siemens Healthcare, Forchheim, Germany) equipped with an integrated high-resolution detector (Stellar Technology, Siemens). 
Intravenous contrast (Omnipaque, GE Healthcare, Princeton, NJ) was administered using power injector (\SI{5}{\milli\liter\per\second}) via the ante-cubital vein followed by saline flush (60--90 \si{\milli\liter} total contrast volume).
Descending aorta contrast-triggered (100 Hounsfield units [HU] at 120kVp), electrocardiogram (ECG)–gated formal CT data acquisition was begun on reaching this threshold with a 10 second delay. 
CT parameters include a slice collimation of \num{192}$\times$\num{0.6} \si{\milli\meter}, gantry rotation time of \SI{250}{\milli\second}, pitch of \num{3.2}. 
Automated tube current modulation was performed using a reference tube current–time product of \SI{400}{\milli\ampere\second} and using automated attenuation-based tube voltage selection with a reference tube potential of 120 kVp. 
Initial retrospective ECG-gated scans were reconstructed in $\SI{5}{\percent}$ phase increments throughout the cardiac cycle using iterative reconstruction, slice thickness of \SI{0.6}{\milli\meter} and an increment of \SI{0.4}{\milli\meter}.
Patient data is summarized in \autoref{tab:patient_data}.

\begin{table}
\caption{Patient data.}
\label{tab:patient_data}
\centering
\begin{tabular}{|l|c|}
\hline
Parameter & 
Value \\ 
\hline
Left ventricular ejection fraction (LVEF) & $\SI{34}{\percent}$\\
Left ventricular end diastolic volume (LVEDV) & $\SI{414}{\milli\liter}$\\
Left ventricular end systolic volume (LVESV) & $\SI{274}{\milli\liter}$\\
Hear rate (HR) & 83 bpm \\
Cardiac output (CO) & $\SI{11.62}{\liter\per\minute}$ \\
Systolic cuff pressure ($\text{P}_\text{sys}^\text{cuff}$) & $\SI{97}{\mmHg}$\\
Diastolic cuff pressure ($\text{P}_\text{dia}^\text{cuff}$) & $\SI{57}{\mmHg}$\\
Gender & male \\
Age & \num{74} \\
\hline
\end{tabular}
\end{table}
\subsection{Model Generation}\label{sec:model_gen}
\begin{figure*}[!t]
\centerline{\includegraphics[width=7.16in]{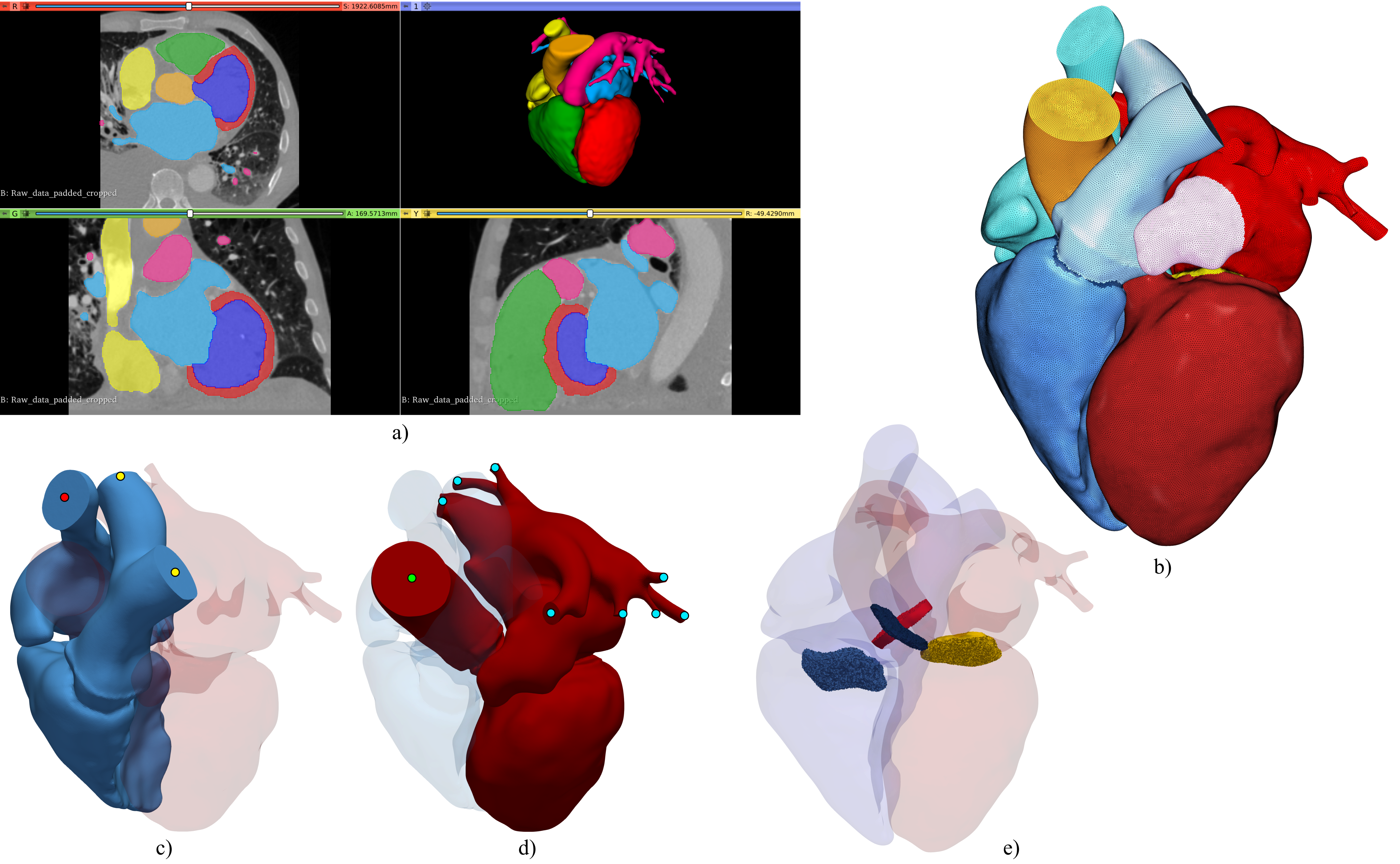}}
\caption{Whole heart model generation workflow. Shown are a) the pre-final segmentation in \texttt{Slicer} before automatically adding valve regions, 
b) the final multi-label mesh, 
the outflow boundaries for c) the right heart, marked by red and yellow circles, 
and d) the left heart, marked by green and blue circles, and 
e) the automatically generated valve regions.}
\label{fig:sim_setup}
\end{figure*}

Cardiac anatomy was automatically segmented from the CT DICOM images \cite{Zheng2008,Strocchi2020,Rodero2021}, to provide labels for all cardiac chambers and major vessels (\autoref{fig:sim_setup}a). 
Additional post processing was performed using \texttt{Seg3D}\footnote{\url{https://www.sci.utah.edu/cibc-software/seg3d.html}} and \texttt{Slicer}\footnote{\url{https://www.slicer.org/}}
to obtain 16 labels comprising left ventricle (LV), right ventricle (RV), left atrium (LA), right atrium (RA), aorta (AO), and pulmonary artery (PA) blood pools as well as labels encoding the locations of aortic valve (AV), mitral valve (MV), pulmonary valve (PV) and tricuspid valve (TV).
Valve labels were automatically generated as thin voxel regions between compartment regions see \autoref{fig:sim_setup}e).
Multilabel segmentations were used to create an unstructured finite element surface mesh using \texttt{CGAL}\footnote{\url{https://www.cgal.org/}},
which served as input for the unstructured volumetric mesh generation, 
including three prismatic boundary layers, using the software package \texttt{Meshtool} \cite{Neic2020} (see \autoref{fig:sim_setup}b).
Cardiac kinematics was extracted over one cardiac cycle by non-rigid registration, 
using the sparse free-form deformation (SFFD) technique \cite{Shi2013} 
that extends the classic FFD approach and recovers smoother displacement fields \cite{Razeghi2020a, Razeghi2020b}.
\subsection{Computational Methods \& Simulation}\label{sec:comp_meth}
\begin{figure}[!t]
\centerline{\includegraphics[width=\columnwidth]{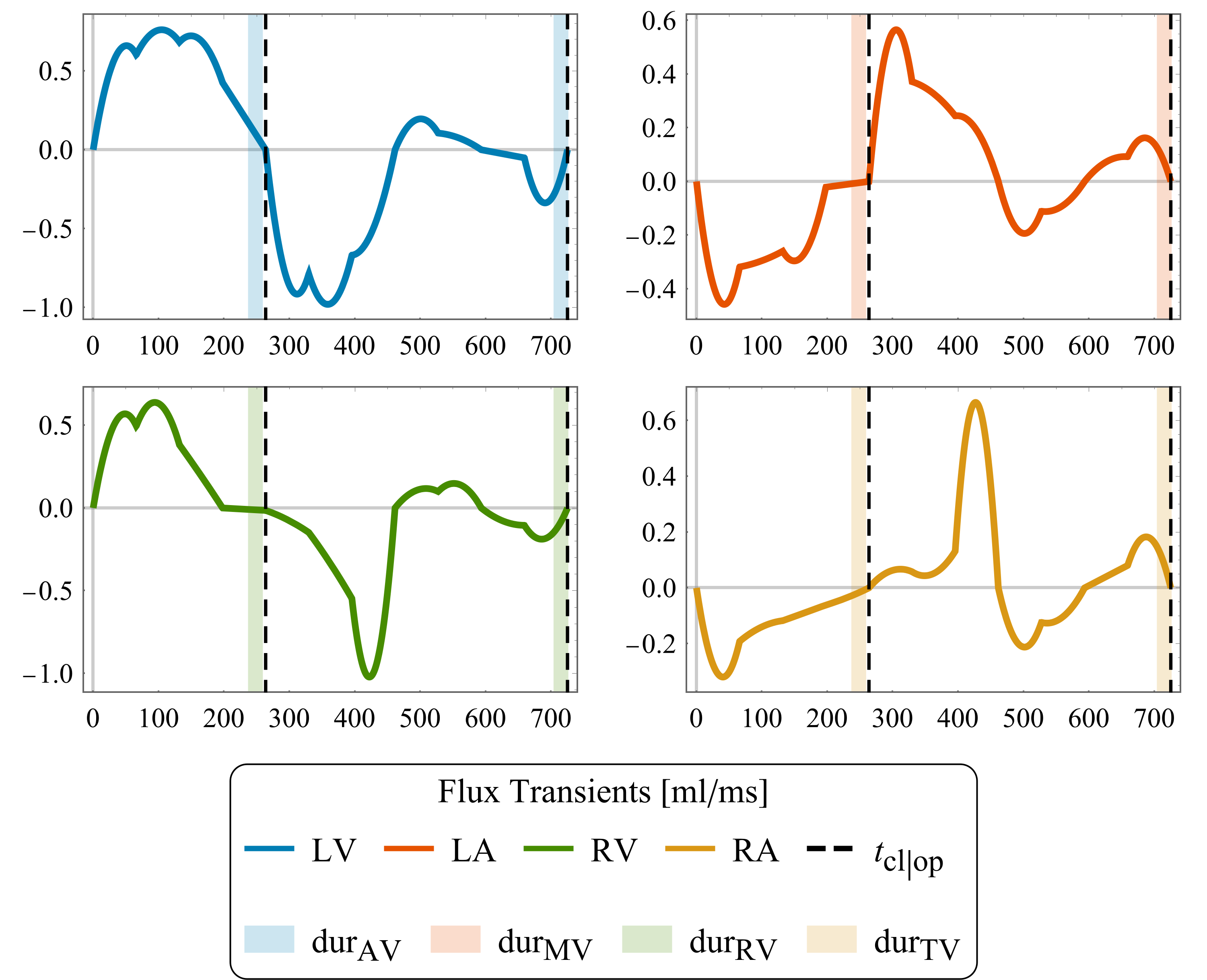}}
\caption{Fluxes computed from volume transients of \autoref{fig:volume_transients}. Dashed lines indicated timings of valves switching, with opaque bars indicating 
the duration of switching.}
\label{fig:flux_transients}
\end{figure}

\begin{figure}[!t]
\centerline{\includegraphics[width=\columnwidth]{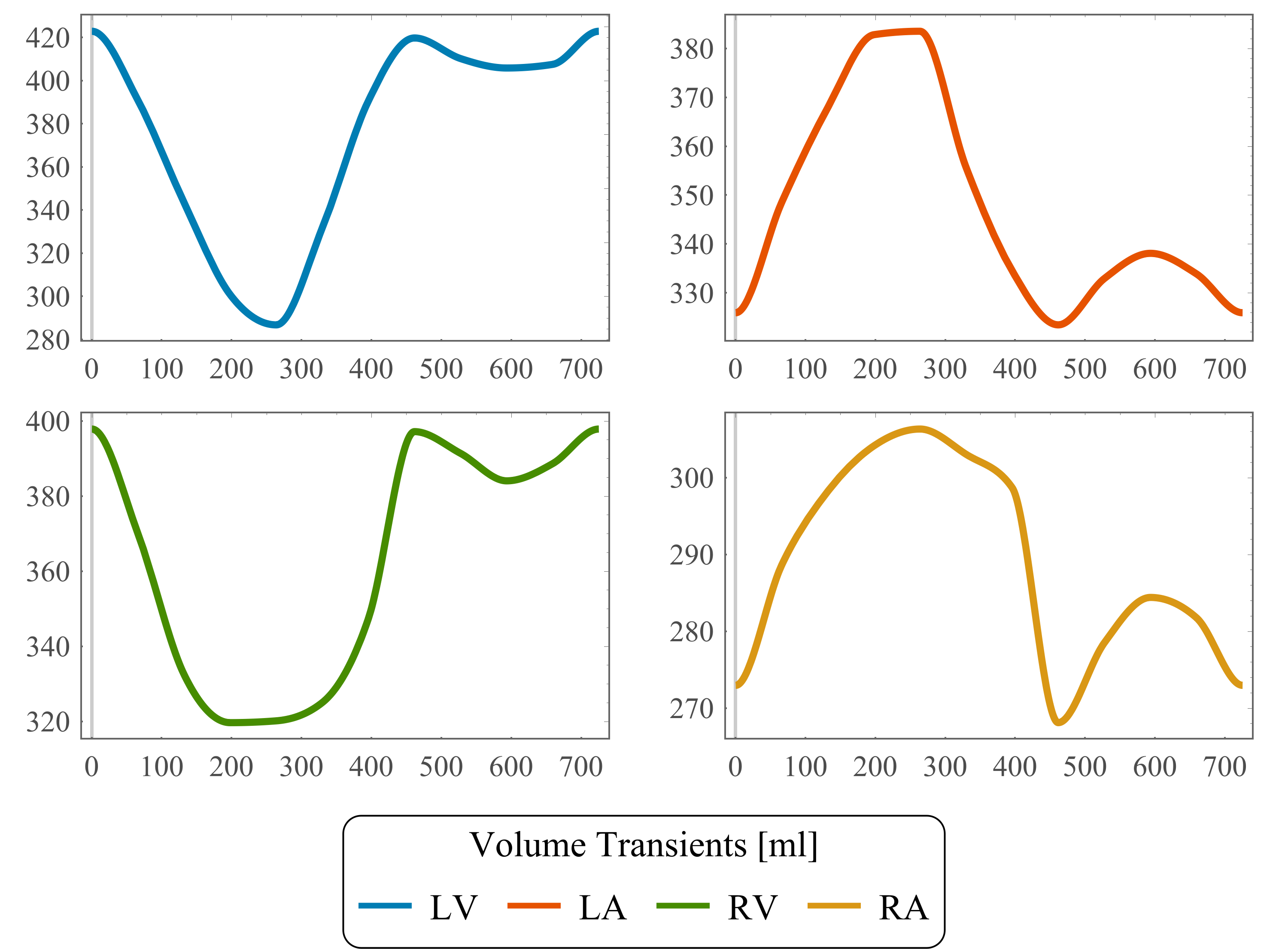}}
\caption{Volume transients extracted from the registered mesh motion 
of LV, RV, LA and RA blood pool.}
\label{fig:volume_transients}
\end{figure}

Image derived kinematics was used as input to drive the CFD model of whole-heart hemodynamics.
With prescribed motion, blood flow in the left and right heart 
can be simulated independently.
Assuming Newtonian blood flow, hemodynamics is modeled with an arbitrary Lagrangian-Eulerian (ALE) formulation of the Navier-Stokes equations \cite{Hughes1981, LeTallec2001}. 
The effect of heart valves upon blood flow is taken into account
by including an ad-hoc extension to the ALE-Navier-Stokes-Brinkman (ALE-NSB) equations
with an added Darcy drag term penalizing flow in the areas covered by the valves  \cite{Fuchsberger2021, Engels2016, Angot1999feb, Daub2020}.
Extensions required for moving domains are explained in more detail in Supplement \ref{S-supplalensb}.
Computational domains labeled as valves are parameterized by a penalty parameter $\kappa_*$, modeling vanishing permeability, with $*$ denoting any of the four heart valves, AV, MV, PV, TV, and the duration $\mathrm{dur}_*$ (see \autoref{fig:flux_transients} for an illustration) of valve opening and closing.
A RBVMS discretization is used\cite{Bazilevs2007}, adapted to the ALE-NSB equations.
A generalized-$\alpha$ integrator \cite{Jansen2000} with $\rho_\infty = \num{0.2}$ is employed for time discretization 
and the arising non-linear systems are solved with an inexact Newton-Raphson method \cite{Dembo1982}.
Domain motion was extended into the interior of the blood pool 
using a linear elastic model optimized for retaining finite element quality.
Dirichlet displacement boundary conditions are used at the blood pool walls 
enforcing a velocity matching the time derivative of the registered cardiac motion.
On the arterial outlets (aorta and pulmonary artery) we used $0D$ three element Windkessel models \cite{FouchetIncaux2014}.
Windkessel parameters of systemic circulation comprising characteristic impedance, $\mathrm Z_\mathrm{WK}$, 
resistance $\mathrm{R}_\mathrm{WK}$ and compliance $\mathrm{C}_\mathrm{WK}$ 
were determined from cuff pressure measurements \cite{Karabelas2018,Marx2020}.
This resulted in $\mathrm{R}_\mathrm{WK}=\SI{49.89}{\kilo\pascal\milli\second\per\milli\liter}$.
Values for $\mathrm{Z}_\mathrm{WK}$ and $\mathrm{C}_\mathrm{WK}$ were determined as $\num{0.05} \mathrm{R}_\mathrm{WK}$ and $\mathrm{C}_\mathrm{WK}=\frac{\mathrm{HR}}{\mathrm{R}_\mathrm{WK}}$ respectively.
As no pressure measurements were available for the right heart, 
Windkessel parameters for the pulmonary circulation were estimated 
by assuming a default value of $\SI{14}{\mmHg}$ 
for mean pulmonary artery pressure \cite{Kovacs2009} and estimating Windkessel parameters from this value.
RV cardiac output was estimated from its end diastolic and end systolic volume, 
with the latter estimated from the volume transients in \autoref{fig:volume_transients}.
At the other outlets pressures $p_\mathrm{LA}=\SI{10}{\mmHg}$ and $p_\mathrm{RA}=\SI{5}{\mmHg}$ were prescribed. 
The location of all outlets are illustrated in \autoref{fig:sim_setup}c) and \autoref{fig:sim_setup}d).
For numerical stability the directional do-nothing outflow stabilization \cite{Braack2014} was used.

\subsection{Global Sensitivity Analysis}\label{sec:GSA_meth}
To quantify the impact of input parameters on the total variance of output features 
global sensitivity analysis (GSA) using \emph{Gaussian process emulation} (GPE) 
was employed to replace the highly non-linear computationally expensive map 
from parameters to features with a fast-evaluating, probabilistic surrogate map.
We selected $D$ parameters and $M$ characteristic output features for the studied model.
GPEs were trained following \cite{Longobardi2020}. 
Briefly, we used a $\approx 10D$ sized sample drawn from a Latin hypercube design 
in the $D$-dimensional parameter space with initial ranges given by $\SI{\pm 25}{\percent}$ perturbation around the baseline values.
Model simulations were carried out for each of these parameter sets 
and the successfully completed simulations were collected to build the training dataset.
Simulations where CFD simulation failed to converge were discarded.
GPEs were defined as the sum of a deterministic mean function and a stochastic process \cite{OHagan2006} while the stochastic process is a centered zero-mean Gaussian process with stationary Mat\'ern covariance function \cite{Genton2002}.
The model likelihood was taken to be Gaussian, 
i.e.\ the learning sample observations are modeled to be affected by an additive, independent and identically distributed noise.

\subsection{Computational Framework}
\subsubsection{Computational Fluid Dynamics}
The discretized and linearized block system of the ALE-NSB equations 
was solved for each Newton–Raphson iteration and every time step.
A flexible generalized minimal residual method (fGMRES) and efficient preconditioning based on the libraries PETSc\footnote{\url{https://www.mcs.anl.gov/petsc/}} 
and hypre/BoomerAMG\footnote{\url{https://hypre.readthedocs.io/en/latest/index.html}}
were employed.
CFD model and calculation of residence times have been implemented 
in an extension of the Cardiac Arrhythmia Research Package (\texttt{carpentry}) \cite{Vigmond2008}.
Parallel performance and scalability of \texttt{carpentry} has been previously investigated in \cite{Augustin2016,Karabelas2018}.
Details on numerical aspects are provided in Supplement \ref{S-supplnumsol}.

\subsubsection{GPE Training}
All the GPE's (hyper)parameters
were jointly optimized by maximization of the model log-marginal likelihood using \texttt{GPErks} emulation tool\footnote{\url{https://github.com/stelong/GPErks}} based on the \texttt{GPyTorch} Python library\footnote{\url{https://gpytorch.ai/}}.
Univariate GPEs were trained to predict each output feature using a $\num{5}$-fold cross-validation process.
GPEs' accuracy was evaluated using the average $R^2$-score across the obtained scores when testing the emulators on the respective left-out parts of the dataset.
The so trained GPEs were used as emulators for the global sensitivity analysis.
Model outputs' sensitivity to parameters was characterized by Sobol' first-order $S_1$ and total effects $S_T$~\cite{Kucherenko2005}.

\subsection{Data Analysis}
\begin{figure}[!t]
\centerline{\includegraphics[width=\columnwidth]{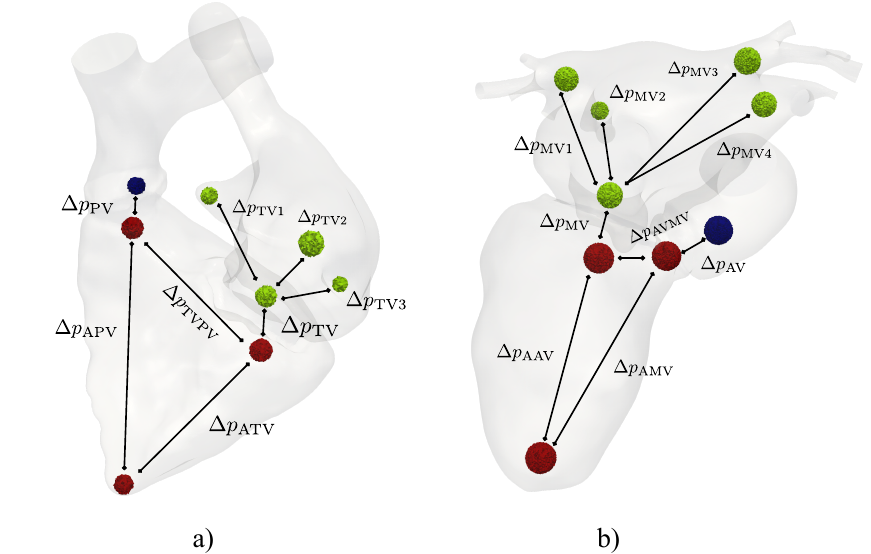}}
\caption{Illustration of areas in the left and right heart used to compute pressure drops and differences respectively. A black line denotes that the pressure difference between those areas is calculated.}
\label{fig:sphere_setup}
\end{figure}

Pressure gradients and differences as well as flow velocities 
were calculated by computing spatial averages over spherical regions 
chosen as observation sites, see \autoref{fig:sphere_setup}b.
Output features used for training were calculated from derived quantities by temporal averaging, or taking the temporal maximum over the whole cardiac cycle.

\section{Results}
\label{sec:results}

\subsection{Simulation}
\begin{figure*}[!t]
\centerline{\includegraphics[width=7.16in]{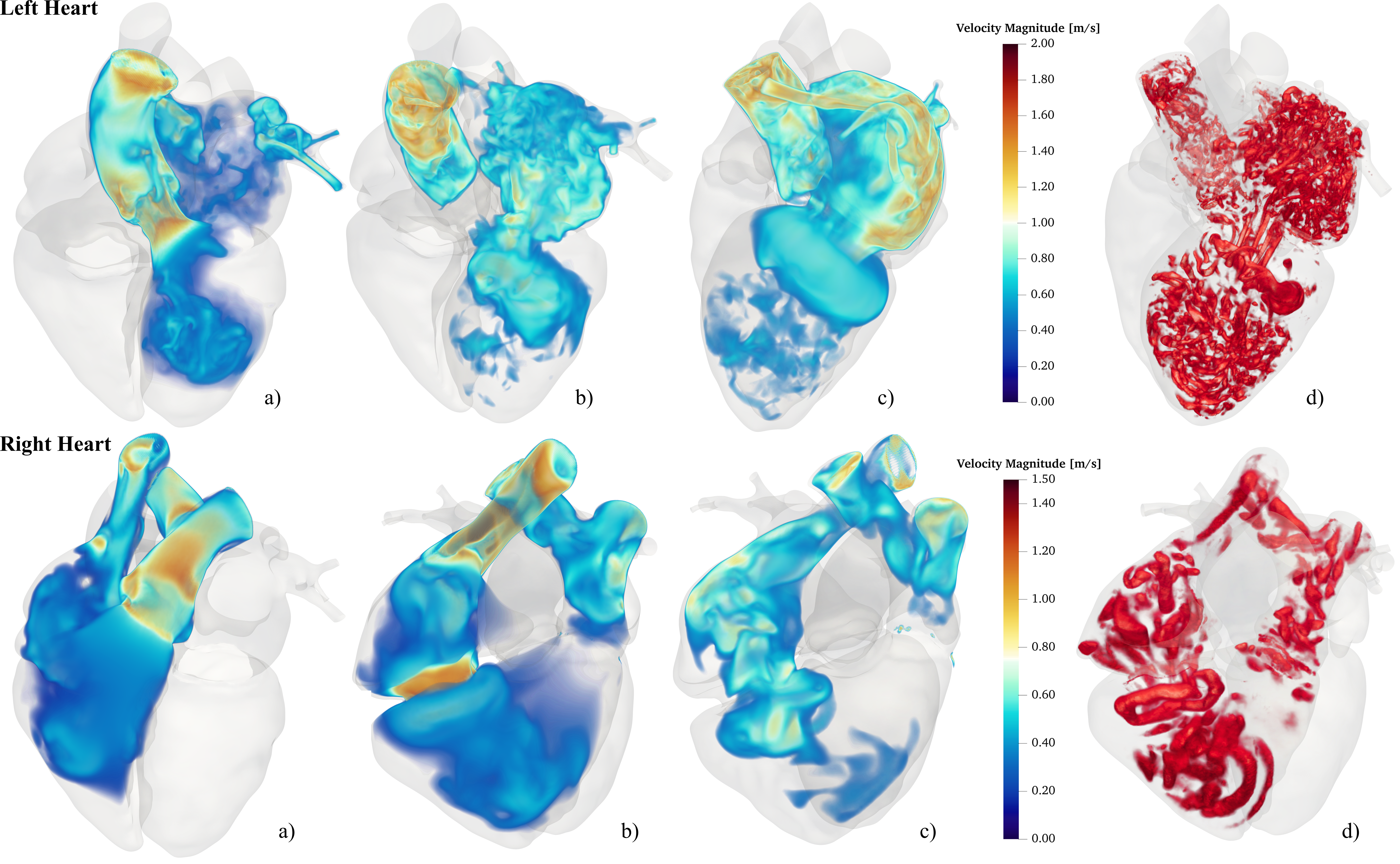}}
\caption{
CFD results show  left (top panels) and right (bottom panels) heart blood flow at 
a) peak systole, b) end of systole, and c) peak diastole,
and d) the strain normalized $Q$ criterion at peak diastole.
}
\label{fig:cfd_results}
\end{figure*}

Four heartbeats were simulated at a time step of $\Delta t = \SI{0.3625}{\milli\second}$ resulting in $16000$ time steps.
Simulations were carried out on the \emph{Vienna Scientific Cluster 4} (VSC4) 
using 1152 MPI processes and 672 MPI processes,
with an average run time per time step of $\SI{18}{\second}$ and $\SI{9}{\second}$ 
and a total run time of $\SI{80}{\hour}$ and $\SI{40}{\hour}$ 
for left and right heart simulations, respectively.
Volume renderings of the velocity magnitude at various time instances are shown in
\autoref{fig:cfd_results}.
The large-scale flow characteristics in both ventricles is the formation of an asymmetric vortex ring (\autoref{fig:cfd_results}c) and \autoref{fig:cfd_results}e)) 
next to the MV and TV traveling towards the apex,
also apparent in the visualization of the strain-normalized $Q$ criterion 
in \autoref{fig:cfd_results}d) and \autoref{fig:cfd_results}h).
As expected, jet formation is witnessed at the opening of the heart valves, see rightmost subfigures of \autoref{fig:cfd_results}.
Furthermore, flow in the AO shows strong non-laminar behavior and increased flow speeds can be observed in the upper areas of the LA.
A video showing the final heartbeat is available as supplement.

\subsection{Global Sensitivity Analysis Using Surrogate Models}\label{sec:GSA_res}
We performed a GSA for both sides of the heart as outlined in \autoref{sec:GSA_meth}.
First, we used $D=\num{6}$ parameters (summarized in \autoref{tab:input_parameters_gpe}) as key regulators of our left heart model, and we characterized the model behavior at a specific set of parameters using $M=\num{16}$ features with notation and baseline values summarized in \autoref{tab:output_parameters_gpe}.
More specific, we used the following output features:
mean systolic pressure gradient over aortic valve (AV), $\Delta \mathrm{p}_\mathrm{AV}$, and  mean diastolic pressure gradient over mitral valve (MV), $\Delta \mathrm{p}_\mathrm{MV}$, as defined in \cite{Johnson2018}; 
mean pressure difference between four landmark points 
in the LA and MV, $\Delta \mathrm{p}_{\mathrm{MV}1,2,3,4}$;
mean pressure difference between apex and MV, $\Delta \mathrm{p}_{\mathrm{AMV}}$;
mean pressure difference between apex and AV, $\Delta \mathrm{p}_{\mathrm{AAV}}$;
mean pressure difference between AV and MV, $\Delta \mathrm{p}_{\mathrm{AVMV}}$;
mean pressure gradient over PV, $\Delta \mathrm{p}_{\mathrm{PV}}$;
mean pressure gradient over TV, $\Delta \mathrm{p}_{\mathrm{TV}}$;
mean pressure difference between four landmark points in the RA and TV, $\Delta \mathrm{p}_{\mathrm{TV}1,2,3}$;
mean kinetic energy LV, $\mathrm{E}_{\mathrm{k},\mathrm{LV}}$;
mean kinetic energy AO, $\mathrm{E}_{\mathrm{k},\mathrm{AO}}$;
mean kinetic energy LA, $\mathrm{E}_{\mathrm{k},\mathrm{LA}}$;
mean kinetic energy RV, $\mathrm{E}_{\mathrm{k},\mathrm{RV}}$;
mean kinetic energy PV, $\mathrm{E}_{\mathrm{k},\mathrm{PV}}$; 
mean kinetic energy RA, $\mathrm{E}_{\mathrm{k},\mathrm{RA}}$;
average residence time, LV $\mathrm{RT}_\mathrm{LV}$;
average residence time, RV $\mathrm{RT}_\mathrm{RV}$;
average residence time, left atrial appendage (LAAPP) $\mathrm{RT}_\mathrm{APP}$;
maximal velocity magnitude over AV, $\mathrm{maxv}_{\mathrm{AV}}$,
MV, $\mathrm{maxv}_{\mathrm{MV}}$, PV, $\mathrm{maxv}_{\mathrm{PV}}$, and TV, $\mathrm{maxv}_{\mathrm{TV}}$.
Residence times were calculated using an continuum approach described in \cite{Long2013} solved with the flux corrected transport finite element method (FCT-FEM) \cite{John2008} adapted to moving grids, see Supplement \ref{S-supplrrt}.

\begin{figure*}[!t]
\centerline{\includegraphics[width=7.16in]{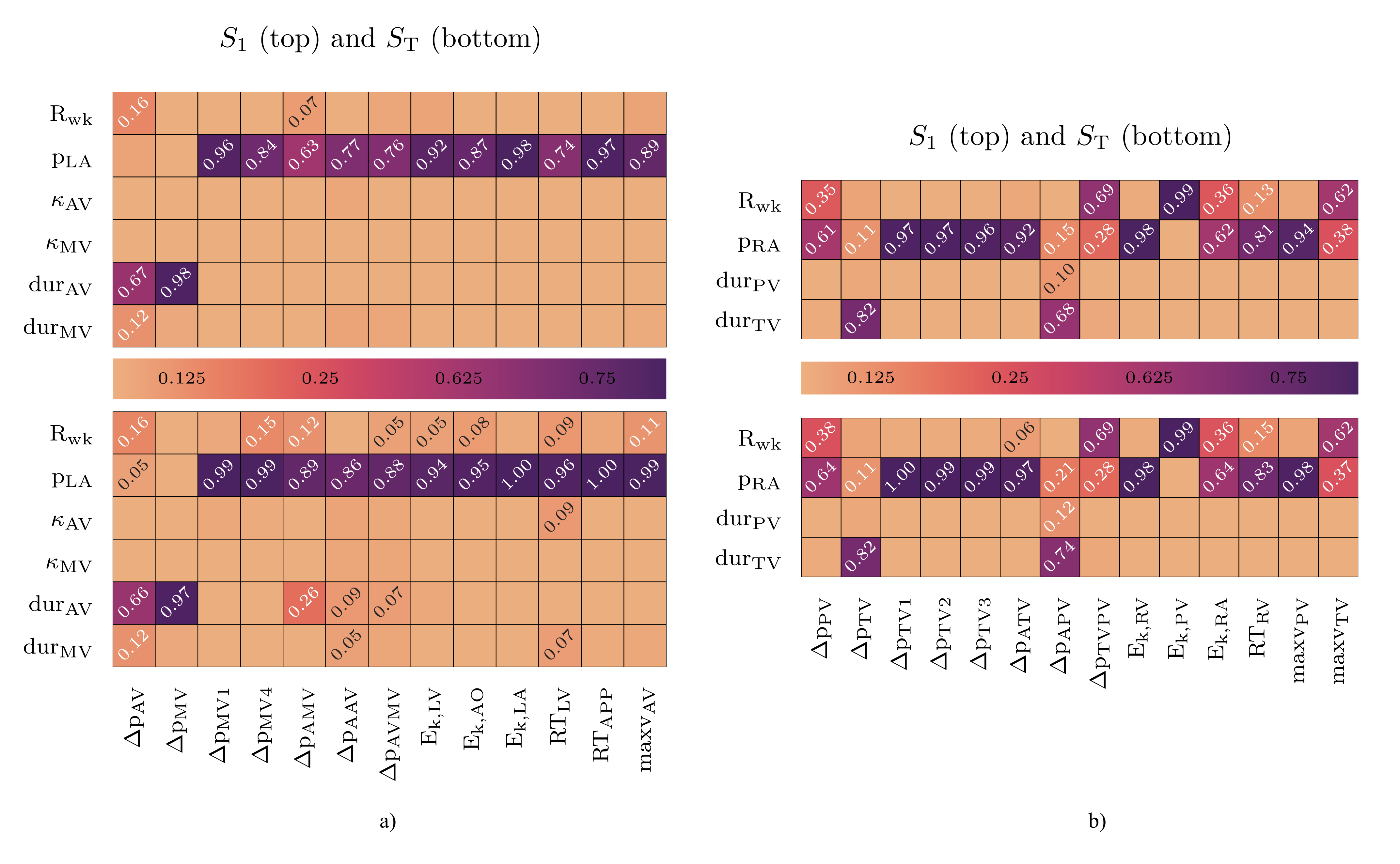}}
\caption{Heat maps of first and total order Sobol indices for the a) left heart and b) right heart GSA.}
\label{fig:LH_RH_GSA}
\end{figure*}

As described in \autoref{sec:GSA_meth}, we used \texttt{GPErks} to incorporate full GPE's posterior distribution samples to estimate the first and total Sobol' indices $S_1$ and $S_T$ using Saltelli's method \cite{Saltelli2010} with $n=\num{10000}$ samples drawn from each GPE.
Sobol indices were calculated with the help of \texttt{SALib} python library \cite{Herman2017}.
Only GPEs having a mean $R^2$ test score $>0.5$ were used for indices calculation. 
This resulted in excluding features $\mathrm{maxv}_\mathrm{MV}$, $\Delta p_\mathrm{MV2}$, $\Delta p_\mathrm{MV3}$, and $\Delta p_\mathrm{AVMV}$ from the analysis. 
Parameters with resulting indices below $\num{0.01}$ were considered to have no/negligible effect.
The resulting indices are summarized as heat-maps in \autoref{fig:LH_RH_GSA}a).
From GSA we concluded that the penalization parameters $\kappa_\mathrm{AV}$, and $\kappa_\mathrm{MV}$ have no or negligible effect and feature $p_\mathrm{LA}$ has a strong effect.
The same procedure was carried out for the right bloodpool model with penalization parameters $\kappa_*$ removed from the training phase due to negligible influence.
We chose similar output features summarized in \autoref{tab:output_parameters_gpe}.
Results are summarized in \autoref{fig:LH_RH_GSA}b) showing a strong effect of $p_\mathrm{RA}$ and $R_\mathrm{WK}$.

\begin{table}
\caption{Parameters identified for GPE training.}
\label{tab:input_parameters_gpe}
\centering
\begin{tabular}{|c|c|p{0.9in}|}
\hline
Parameter & 
Range &
Description \\ 
\hline
$R_{\mathrm{WK},\mathrm{AO}}$ &
\SIrange[per-mode=fraction]{37.46}{62.32}{\kilo\pascal\milli\second\per\milli\liter} &
AO Windkessel resistance\\
$R_{\mathrm{WK},\mathrm{PA}}$ &
\SIrange[per-mode=fraction]{27.81}{46.21}{\kilo\pascal\milli\second\per\milli\liter} &
PA Windkessel resistance\\
$p_\mathrm{LA}$ &
\SIrange{7.5}{12.5}{\mmHg} &
LA outlet pressure \\
$p_\mathrm{RA}$ &
\SIrange{3.5}{8.5}{\mmHg} &
RA outlet pressure \\
$\kappa_\mathrm{AV}$ &
\numrange{1e-5}{1e-9} &
AV penalization parameter \\
$\kappa_\mathrm{MV}$ &
\numrange{1e-5}{1e-9} &
MV penalization parameter \\
$\mathrm{dur}_\mathrm{*}$ &
\SIrange{11.25}{18.75}{\milli\second} &
Valve transition times
with $* \in \{\mathrm{AV}, \mathrm{MV}, \mathrm{PV}, \mathrm{TV}\}$\\
\hline
\end{tabular}
\end{table}

\begin{table}
\caption{Output features for GPE training with reference values 
extracted from CFD simulations. 
Reported are temporal means, except for velocities reported as temporal maxima. Clinical measurements if reported are given as means of three measurements.
}
\label{tab:output_parameters_gpe}
\centering
\begin{tabular}{|l|c|c|}
\hline
Parameter &
\emph{in silico} Reference Value &
Clinical Measurements
\\ 
\hline
$\Delta \mathrm{p}_\mathrm{AV}$ &
\SI{4.61}{\mmHg} & \SI{5.0}{\mmHg}\\
$\Delta \mathrm{p}_\mathrm{MV}$ &
\SI{2.71}{\mmHg} & \SI{2.38}{\mmHg}\\
$\Delta \mathrm{p}_{\mathrm{MV}1,2,3,4}$ &
\bigcell{c}{\num{-0.0106} \\ \num{0.225} \\ \num{0.183} \\ \num{-0.0025}} \si{\mmHg} & \\
$\Delta \mathrm{p}_{\mathrm{AMV}}$ &
\SI{1.732}{\mmHg} & \\
$\Delta \mathrm{p}_{\mathrm{AAV}}$ &
\SI{1.60}{\mmHg} & \\
$\Delta \mathrm{p}_{\mathrm{AVMV}}$ &
\SI{0.21}{\mmHg} & \\
$\Delta \mathrm{p}_{\mathrm{PV}}$ &
\SI{2.35}{\mmHg} & \SI{3.0}{\mmHg}\\
$\Delta \mathrm{p}_{\mathrm{TV}}$ &
\SI{5.73}{\mmHg} & \SI{47.0}{\mmHg}\\
$\Delta \mathrm{p}_{\mathrm{TV}1,2,3}$ &
\bigcell{c}{\num{0.204} \\ \num{0.342} \\ \num{0.228}} \si{\mmHg} & \\
$\mathrm{E}_{\mathrm{k},\mathrm{LV}}$ &
\SI{16.71}{\milli\joule} & \\
$\mathrm{E}_{\mathrm{k},\mathrm{AO}}$ &
\SI{22.59}{\milli\joule} & \\
$\mathrm{E}_{\mathrm{k},\mathrm{LA}}$ &
\SI{23.33}{\milli\joule} & \\
$\mathrm{E}_{\mathrm{k},\mathrm{RV}}$ &
\SI{3.65}{\milli\joule} & \\
$\mathrm{E}_{\mathrm{k},\mathrm{PV}}$ &
\SI{5.51}{\milli\joule} & \\
$\mathrm{E}_{\mathrm{k},\mathrm{RA}}$ &
\SI{9.59}{\milli\joule} & \\
$\mathrm{RT}_\mathrm{LV}$ &
\SI{0.811}{\second} & \\
$\mathrm{RT}_\mathrm{APP}$ &
\SI{0.854}{\second} & \\
$\mathrm{RT}_\mathrm{RV}$ &
\SI{0.91}{\second} & \\
$\mathrm{maxv}_{\mathrm{AV}}$ &
\SI{1.13}{\meter\per\second} & \SI{1.15}{\meter\per\second}\\
$\mathrm{maxv}_{\mathrm{MV}}$ & 
\SI{0.73}{\meter\per\second} & \SI{0.81}{\meter\per\second}\\
$\mathrm{maxv}_{\mathrm{PV}}$ &
\SI{0.71}{\meter\per\second} &  \SI{0.814}{\meter\per\second}\\
$\mathrm{maxv}_{\mathrm{TV}}$ &
\SI{0.57}{\meter\per\second} & \SI{3.43}{\meter\per\second}\\
\hline
\end{tabular}
\end{table}

\section{Discussion}
\label{sec:disscussion}
Being able to identify key parameters and regulators in a hemodynamic CFD model of the human heart is paramount for personalization.
However, personalization of four chamber CFD models is computationally expensive.
Here we show that the use of ALE-NSB allows computationally tractable simulations, the GPE can be used to emulate most outputs, and that pre load is the key parameter in determining boundary driven four chamber heart CFD models. 
Our CFD simulations took between \SI{10}{\hour} and \SI{20}{\hour} per heart beat for the left or right side of the heart.
This breaks down to an average wallclock time of $\approx$ \SI{11}{\second} for performing one nonlinear time step of the CFD simulator.
Comparing our average wallclock times with other approaches, for instance \SI{11}{\second} reported in \cite{Griffith2009} using IBM, or \SI{30}{\second} - \SI{50}{\second} reported in \cite{Forti2015} using a semi-implicit algorithm with higher order finite elements, or \SI{50}{\second} reported in \cite{Daub2018} using a similar algorithm as in this manuscript, we find that our ALE-NSB method provides a competitive implementation putting us well into the forecasted optimal wallclock times for hemodynamic CFD simulations shown in \cite{Mittal2016}.

Comparing with clinically measured data in \autoref{tab:output_parameters_gpe} we 
saw good agreement for $\mathrm{maxv}_\mathrm{AV,MV,PV}$ with relative error of $\approx$ \SI{2}{\percent}, \SI{10}{\percent}, and \SI{13}{\percent} similarly for $\Delta p_\mathrm{AV,MV,PV}$ with relative error of $\approx$ \SI{8}{\percent}, \SI{13}{\percent}, and \SI{24}{\percent}.
Clinical data suggested a possible TV regurgitation.
We did not aim to capture TV regurgitation in the simulations, and this likely explains the discrepancy in $\mathrm{maxv}_\mathrm{TV}$ and $\Delta p_\mathrm{TV}$.

There is growing interest in using reduced order models and physics informed neural networks (PINNs) for accelerating or creating model surrogates \cite{Sun2020}.
Each method has its purpose, here we show that GPEs, which are both fast and provide an estimate of uncertainty can be used to emulate most, but not all, four chamber heart CFD simulation outputs using $\approx$ 10 simulations per parameter.
To train our GPEs we used in total \num{180} CFD simulations comprising 4 heart beats each.
Executing those simulations took approximately \SI{7700}{\hour} of wallclock time on the HPC clusters VSC4 (AT) and ARCHER (UK).
Using those data sets we performed the first GSA of model parameters for informing cavity driven flow.
Training of the GPEs and running GSA took approximately \SI{5}{\hour}.
Running GSA without a surrogate model would have resulted in intractable amounts of CFD simulations highlighting the possible savings in computing time and resources.

Output features $\mathrm{maxv}_\mathrm{MV}$, $\Delta p_{\mathrm{MV}2}$, and $\Delta p_{\mathrm{MV}3}$ showed $R^2$ test scores below \num{0.5}.
As the GPEs are based on nonlinear CFD simulations, it is hard to give a definite answer as to why those particular features were excluded.
Possible explanations could be, underresolved CFD grids close to extraction points of the features, or lacking temporal resolution.

We identified the pre-load as a key variable in defining simulation clinical outputs, in both the atrial and ventricle flow in four chamber boundary driven flow simulations. This highlights the need to have an accurate estimate of pre-load when performing  boundary driven CFD simulations. 
As blood flows from the atria to the ventricle and then out through the aorta (or pulmonary artery) the parameters that impact atrial flow will impact down stream flows. 
Conversely, the after-load properties only impact blood flow out of the ventricle and do not directly impact the atrial flow. 
This potentially explains the importance of pre-load over after-load in our simulations.
Furthermore, we considered time averages over the complete heart beat.
During systole, pressure signals are not sensitive to any of the input parameters.
However, this changes in diastole and we provide an additional explanation in Supplement \ref{S-supplinterp}.

It is important to note that our findings are for the specific case of boundary driven flow and do not reflect the relative importance of pre-load and after-load in patients, where after-load can feedback on ventricle function, and hence atrial filling, so may play a far greater role physiologically.

\section{Conclusion}
\label{sec:conclusions}
In this work we presented full GSA based on GPE surrogate models for four chamber heart hemodynamics.
We showed that modeling valves using a penalization approach is independent of numerical parameters.
GSA revealed strong influences of left and right atrial pressure and medium influence of arterial and pulmonary arterial resistances.
These results show the possibility and potential speedup using surrogate models to replace full-blown CFD models for human heart hemodynamics.

\bibliographystyle{IEEEtran}
\bibliography{ms}   %
\makeatletter\@input{xx.tex}\makeatother
\end{document}


\title{Supplementary Material for \enquote{Global Sensitivity Analysis of Four Chamber Heart Hemodynamics Using Surrogate Models}}
\author{E.~Karabelas, S.~Longobardi, J.~Fuchsberger, O.~Razeghi, C.~Rodero,\\R.~Rajani, M.~Strocchi, G.~Haase, G.~Plank, and S.~Niederer}
\date{\today}
\maketitle

\section{Residual-Based Variational Multiscale Formulation for Navier-Stokes-Brinkman Equations on moving domains}
\label{supplalensb}
The Navier-Stokes-Brinkman (NSB) equations, originating from porous media theory, can be employed with the purpose of simulating viscous flow including complex shaped solid obstacles in a fluid domain, see \cite{arquis1984conditions}, and \cite{Angot1999feb,carbou2003} for a in-depth mathematical analysis.
The NSB model was successfully extended to moving obstacles and applied to model flapping insect flight in \cite{Engels2016}.
In the present work, we use the NSB equations including the adaptation for moving obstacles as well as moving domains using the \emph{arbitrary lagrangian Eulerian} (ALE) formulation \cite{Hirt1974,Hughes1981, LeTallec2001}:
\begin{align}
    &\rho \left( \frac{\partial}{\partial t} \vec{u} +(\vec{u} - \vec w) \cdot \nabla \vec{u} \right)- \nabla \cdot \tensor{\sigma}(\vec{u},p)+\frac{\mu}{K} (\vec{u}-\vec{u}_s) = 0  \quad  &\mathrm{in} \, \, \mathbb{R}^+ \times \Omega(t), \label{eq:NSB}\\
    &\nabla \cdot \vec{u}=0  &\mathrm{in} \, \, \mathbb{R}^+ \times \Omega(t), \label{eq:incomp}\\
    &\vec{u}=\vec w &\mathrm{on} \, \, \Gamma_{\mathrm{noslip}}(t), \label{eq:noslip}\\
    &\tensor{\sigma}\vec{n}-\rho \beta ((\vec{u} - \vec w)\cdot\vec{n})_-=\vec{h} &\mathrm{on} \, \, \Gamma_{\mathrm{outflow}}(t),\label{eq:outflow}\\
    &\vec{u}=\vec{g} &\mathrm{on} \, \, \Gamma_{\mathrm{inflow}}\label{eq:inflow}(t),\\
    &\left.\vec{u}\right|_{t=0}=\vec{u}_0,
\end{align}
with the time dependent fluid domain $\Omega(t)$ defined as
\begin{align*}
    \Omega(t) := \set{ \vec x \mid \vec x = \vec X + \vec d(\vec X, t), \forall \vec X \in \Omega_0},
\end{align*}
using the ALE mapping $\vec d$ transforming an arbitrary reference configuration $\Omega_0$ into the current fluid domain $\Omega(t)$.
Here $p$, $\vec{u}$, and $\vec w := \tfrac{\partial}{\partial t} \vec d$ represent the fluid pressure, the flow velocity, and the ALE velocity respectively, $\mu$ is the dynamic viscosity and $\rho$ the density.
The volume penalization term $\frac{\mu}{K(t,\vec x)}\vec{u}(t,\vec x)$ is commonly known as \emph{Darcy drag} which is characterized by the permeability $K(t,\vec x)$.
In \eqref{eq:NSB} the Darcy drag is modified to enforce correct no-slip conditions for obstacles moving with the obstacle velocity $\vec{u}_s(\vec{x},t)$. The fluid stress tensor $\tensor{\sigma}(\vec{u},p)$ and strain rate tensor $\tensor{\epsilon}(\vec{u},p)$ are defined as follows:
\begin{align}
    &\tensor{\sigma}(\vec{u},p)=-p \tensor{I}+ 2 \mu \, \tensor{\epsilon}(\vec{u},p),\\
    &\tensor{\epsilon}(\vec{u},p) =\frac{1}{2}\left(\nabla \vec u + \left(\nabla \vec u\right)^\top\right).
\end{align}
For $\vec h=\vec 0$, \eqref{eq:outflow} is known as a directional do-nothing boundary condition \cite{EsmailyMoghadam2011, Braack2014}, where $\vec{n}$ is the outward normal of the fluid domain, $\beta \leq \frac{1}{2}$ is a positive constant and \eqref{eq:backflowstab} is added for backflow stabilization with
\begin{equation}
    ((\vec{u} - \vec w)\cdot\vec{n})_-:=\frac{1}{2}((\vec{u}-\vec w)\cdot\vec{n}-|(\vec{u}-\vec w)\cdot\vec{n}|). \label{eq:backflowstab}
\end{equation}

The ALE domain $\Omega(t)$ is split up into three time dependent sub-domains by means of the permeability $K(t,\vec x)$, namely the fluid sub-domain $\Omega_f(t)$, the porous sub-domain $\Omega_p(t)$ and the solid sub-domain $\Omega_s(t)$.
\begin{equation}
K(t,\vec x)= \label{eq:perm}
\begin{cases}
K_f \rightarrow + \infty & \text{if } \vec x \in \Omega_f(t)  \\
K_p &\text{if } \vec x \in \Omega_p(t)  \\
K_s \rightarrow 0^+ &\text{if } \vec x \in \Omega_s(t)
\end{cases}
\end{equation}

In $\Omega_f(t)$ the classical ALE-Navier–Stokes equations are recovered, while in $\Omega_p$ the full ALE-NSB equations describe fluid flowing trough a moving porous medium, $\vec{u}$ and $p$ are understood in an averaged sense in this context.
In $\Omega_s(t)$ the velocity $\vec{u}$ is approaching $\vec{u}_s$ and thus asymptotically satisfying the no-slip condition on the  $\Omega_f(t) / \Omega_s(t)$ interface.
Note that even in the case where $K \rightarrow 0^+$ the penalization term has a well defined limit, see \citep{Angot1999feb}.

\subsection{Hemodynamic Afterload Models}
Modeling of afterload for hemodynamics is modeled by using a 0D Windkessel model.
This means we define $\vec h$ in \eqref{eq:outflow} as
\begin{align*}
    \vec h := - p_\mathrm{WK}(t) \vec n
\end{align*}
with the \emph{Windkessel pressure} $p_\mathrm{WK}$ is governed by the differential algebraic system \cite{FouchetIncaux2014}
\begin{align}
    C_\mathrm{WK} \frac{\mathrm{d}}{\mathrm{d}t} p_\mathrm{d}(t) + \frac{p_\mathrm{d}(t)}{R_\mathrm{WK}} &= Q(t), \label{eq:wk:1}\\
    p_\mathrm{WK}(t) &= Z_\mathrm{WK} Q(t) + p_\mathrm{d}(t), \label{eq:wk:2}\\
    Q(t) &:= \int\limits_{\Gamma_\mathrm{outflow}} \vec u \cdot \vec n\,\mathrm{d}s_\vec{x}.\label{eq:wk:flux}
\end{align}
In the case of multiple Windkessel outlets we will use the same notation for variables with an added $i$ subscript indicating multiple outlets.
Tools for personalization of the individual Windkessel parameters can be found in \cite{Marx2020}.

\subsection{Variational Formulation and Numerical Stabilization}
\label{sec:varform_ALENSB}
Following \cite{Bazilevs2007,Bazilevs2013} the discrete variational formulation of \eqref{eq:NSB} including the boundary conditions \eqref{eq:outflow}, \eqref{eq:inflow} and \eqref{eq:noslip} can be stated in the following abstract form:
Find $\vec{u}^h \in [\mathcal{S}^1_{h,\vec{g}}(\mathcal{T}_\mathrm{N})]^3, \, p^h \in \mathcal{S}^1_h (\mathcal{T}_\mathrm{N})$ such that, for all $\vec{v}^h \in [\mathcal{S}^1_{h,\vec{0}}(\mathcal{T}_\mathrm{N})]^3 $ and for all $q^h \in \mathcal{S}^1_h(\mathcal{T}_\mathrm{N})$
\begin{equation}
    A_{\mathrm{NS}}(\vec{v}^h,q^h;\vec{u}^h,p^h) + S_\mathrm{RBVMS}(\vec{v}^h,q^h;\vec{u}^h,p^h) = F_{\mathrm{NS}}(\vec v_h)
\end{equation}
with the bilinear form of the NSB equations
\begin{equation}
\begin{aligned}
    &A_{\mathrm{NS}}(\vec{v}^h,q^h;\vec{u}^h,p^h)=\\
    & \int\limits_{\Omega}  \rho \, \vec{v}^h \cdot \left[ \left( \frac{\partial \vec{u}^h}{\partial t}+ (\vec{u}^h -\vec w)\cdot \nabla \vec{u}^h  +  \frac{\nu}{K} (\vec{u}^h -\vec{u}_s^h) \right)+ \tensor{\varepsilon}(\vec{v}^h) : \tensor{\sigma}(\vec{u}^h,p^h) \right]\, \dx \\
    & -\int\limits_{\Gamma_{\mathrm{outflow}}} \rho \beta ((\vec{u}^h-\vec w)\cdot\vec{n})_- \vec{v}^h \cdot \vec{u}^h \, \dsx + \int_{\Omega} q^h \nabla \cdot \vec{u}^h \, \dx  \label{eq:varNSB},
\end{aligned}
\end{equation}
the bilinear form $S_\mathrm{RBVMS}$, which will be explained later in Equation~\eqref{eq:bilinearform:VMS}, and the right hand side contribution
\begin{align}
    F_\mathrm{NS} = -p_\mathrm{WK}\int\limits_{\Gamma_\mathrm{outflow}} \vec n \cdot \vec v_h\, \dsx.
\end{align}
We use standard notation to describe the finite element function space $\mathcal S^1_{h,*}(\mathcal T_N)$ as a conformal trial space of piece-wise linear, globally continuous basis functions $v_h$ over a decomposition $\mathcal T_N$ of $\Omega$ into $N$ finite elements constrained by $v_h=*$ on essential boundaries.
The space $S^1_h(\mathcal T_N)$ denotes the same space without constraints.
For further details we refer to \cite{brenner2007mathematical, steinbach2007numerical}.
As previously described in \cite{Karabelas2018} we utilize the residual based variational multiscale (RBVMS) formulation as proposed in \cite{Bazilevs2007,Bazilevs2013}, providing turbulence modeling in addition to numerical stabilization.
In the following we give a short summary of the changes necessary to use RBVMS methods for the ALE-NSB equations.
Briefly, the RBVMS formulation is based on a decomposition of the solution and weighting function spaces into coarse and fine scale subspaces and the corresponding decomposition of the velocity and the pressure and their respective test functions.
Henceforth the fine scale quantities and their respective test functions shall be denoted with the superscript $'$.
We assume $\vec{u}_s =\vec{u}_s^h$ , quasi-static fine scales ($\frac{\partial \vec{u}'}{\partial t} = 0$), as well as $\frac{\partial \vec{v}^h}{\partial t} = 0$, $\vec{u}' = \vec 0$  on $\partial \Omega(t)$ and incompressibility conditions for $\vec{u}^h$ and $\vec{u}'$.
The fine scale pressure and velocity are approximated in an element-wise manner by means of the residuals $\vec{r}_M$ and $r_C$.
\begin{align}
\vec{u}' &= -\frac{\tau_{\mathrm{SUPS}}}{\rho} \; \vec{r}_M(\vec{u}^h,p^h) \label{app_u}\\
p' &= -\rho \;\nu_{\mathrm{LSIC}}\; r_C(\vec{u}^h)\label{app_p}
\end{align}

The residuals of the NSB equations and the incompressibility constraint are:
\begin{align}
\vec{r}_M(\vec{u}^h,p^h) &= \rho \frac{\partial}{\partial t} \vec{u}^h + \rho (\vec{u}^h -\vec w)\cdot \nabla \vec{u}^h - \nabla \cdot \tensor{\sigma}(\vec{u}^h,p^h)+ \frac{\mu}{K} (\vec{u}^h-\vec{u}^h_s) \label{eq:Mres}\\
r_C(\vec{u}^h) &= \nabla \cdot \vec{u}^h\label{eq:Cres}
\end{align}

Taking all assumptions into consideration and employing the scale decomposition followed by partial integration yields the bilinear form of the RBVMS formulation $S_\mathrm{RBVMS}(\vec{v}^h,q^h;\vec{u}^h,p^h)$,
\begin{equation}
\label{eq:bilinearform:VMS}
\begin{aligned}
& S_\mathrm{RBVMS}(\vec{v}^h,q^h;\vec{u}^h,p^h)=\\
+ &\sum_{\Omega_e \in \mathcal T_N} \int_{\Omega_e} \tau_\mathrm{SUPS}\left( (\vec{u}^h -\vec w)\cdot \nabla \vec{v}^h +  \frac{1}{\rho} \nabla q^h - \frac{\nu}{K} \vec{v}^h\right) \vec{r}_M(\vec{u}^h,p^h) \, \dx  \\
+ &\sum_{\Omega_e \in \mathcal T_N} \int_{\Omega_e} \rho \, \nu_\mathrm{LSIC}\, \nabla \cdot \vec{v}^h \, r_C(\vec{u}^h) \, \dx  \\
- &\sum_{\Omega_e  \in \mathcal T_N} \int_{\Omega_e} \tau_\mathrm{SUPS} \, \vec{v}^h \cdot \left(\vec{r}_M(\vec{u}^h,p^h) \cdot \nabla \vec{u}^h  \right) \dx  \\
- &\sum_{\Omega_e \in \mathcal T_N} \int_{\Omega_e}  \frac{\tau_{\mathrm{SUPS}}^2}{\rho} \nabla  \vec{v}^h : (\vec{r}_M(\vec{u}^h,p^h) \otimes \vec{r}_M(\vec{u}^h,p^h)) \dx.
\end{aligned}
\end{equation}
The residuals \eqref{eq:Mres} and \eqref{eq:Cres} are evaluated for every element $\Omega_e \in \mathcal T_N$.
Following \citep{PhDPauli} the stabilization parameters $\tau_\mathrm{SUPS}$ and $\nu_\mathrm{LSIC}$ are defined as:

\begin{align}
\tau_{\mathrm{SUPS}} &:= \left( \frac{4}{\Delta t ^2} + (\vec{u}^h -\vec w)\cdot \tensor{G} (\vec{u}^h-\vec w) + \left(\frac{\nu}{K}\right)^2+C_I \nu^2 \tensor{G} : \tensor{G} \right)^{-\frac{1}{2}}\\
\nu_{\mathrm{LSIC}} &:= \frac{1}{\text{tr}(\tensor{G}) \, \tau_{\mathrm{SUPS}}}
\end{align}

Here $\tensor{G}$ is the three dimensional element metric tensor defined per finite element as
\begin{align*}
    \tensor G \lvert_{\tau_l} &:= \tensor J_l^{-1} \tensor J_l^{-\top},
\end{align*}
with $\tensor J_l$ being the Jacobian of the transformation of the reference element to the physical finite element $\tau_l \in \mathcal T_N$, $\Delta t$ denotes time step size and $C_I$ is a positive constant, taken as 30, derived from an element-wise inverse estimate.
For further details see \cite{Bazilevs2013,Bazilevs2007}.

\subsection{Numerical Solution Strategy}
\label{supplnumsol}
Spatio-temporal discretization of all PDEs and the solution of the arising systems of equations relied upon the Cardiac Arrhythmia Research Package (CARPentry), see \cite{Vigmond2003}.
For temporal discretization of the ALE-NSB equations we used the generalized-$\alpha$ method, see \cite{Jansen2000} with a spectral radius $\rho_\infty = 0.2$.
For updating the Windkessel pressures $p_\mathrm{WK}$ we discretized \eqref{eq:wk:1} with an implicit Euler method.
For ease of coupling with our CFD solver the calculation of $Q(t)$ in \eqref{eq:wk:flux} is lagged by one Newton iteration.
After discretization in space as described in Section \ref{sec:varform_ALENSB} and temporal discretization using the generalized-$\alpha$ integrator we obtain a nonlinear algebraic system to solve for advancing time from timestep $t_n$ to $t_{n+1}$.
A quasi inexact Newton-Raphson method is used to solve this system with linearization approach similar to \cite{Bazilevs2013} adapted to the NSB equations.
At each iteration a  block system of the form
\begin{align*}
    \begin{bmatrix}
     \tensor K_h & \tensor B_h \\
     \tensor C_h & \tensor D_h
    \end{bmatrix}
    \begin{bmatrix}
      \Delta \vec u \\
      \Delta \vec p
    \end{bmatrix}
    = -
    \begin{bmatrix}
     -\vec{R}_{\mathrm{upper}} \\
     -\vec{R}_{\mathrm{lower}}
    \end{bmatrix},
\end{align*}
is solved
with $\tensor K_h$, $\tensor B_h$, $\tensor C_h$, and $\tensor D_h$ denoting the Jacobian matrices, $\Delta \vec u$, $\Delta \vec p$ representing the velocity and pressure updates and $\vec R_\mathrm{upper}$, $\vec R_\mathrm{lower}$ indicating the residual contributions.
In this regard we use the flexible generalized minimal residual method (fGMRES) and efficient preconditioning based on the \texttt{
PCFIELDSPLIT}\footnote{\url{https://www.mcs.anl.gov/petsc/petsc-current/docs/manualpages/PC/PCFIELDSPLIT.html}} package from the library \emph{PETSc} \cite{petsc-web-page, petsc-user-ref, petsc-efficient} and the incorporated suite \emph{HYPRE BoomerAMG} \cite{Henson2002}.
By extending our previous work \cite{Augustin2016, Karabelas2018, Karabelas2019} we implemented the methods in the finite element code \emph{Cardiac Arrhythmia Research Package} (CARPentry) \cite{Vigmond2003, Vigmond2008}.

\section{Obstacle Representation}
\label{sec:methods:obs_algo}
\begin{figure}[htp]
    \centering
    \begin{picture}(210,180)
    \put(0,0){ \includegraphics[width=9cm]{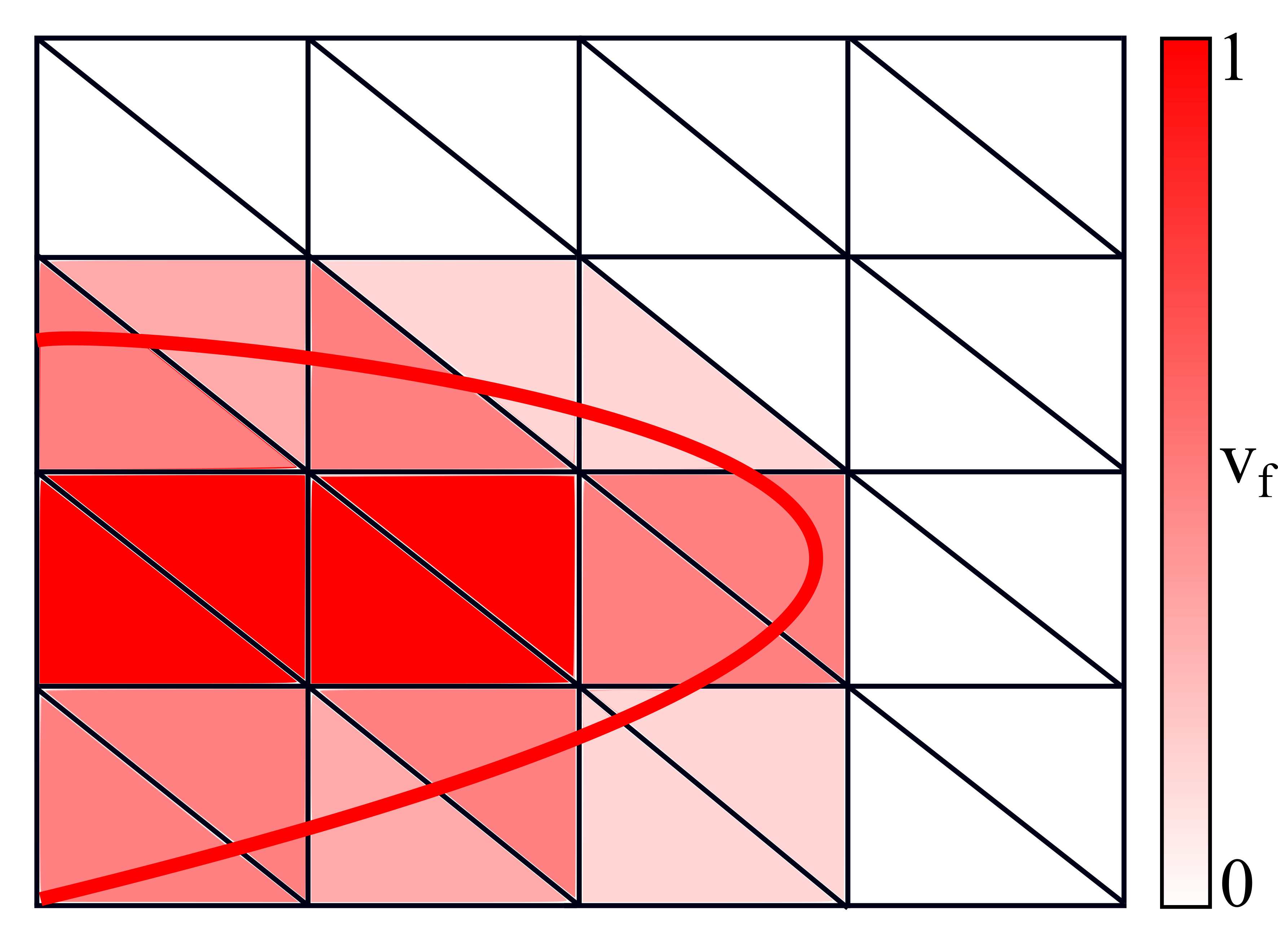}}
    \Large{
    \put(70,60){$\Omega_s$}
    \put(180,150){$\Omega_f$}
    \put(20,107){$\Omega_p$}
    }
    \end{picture}
    \caption{Schematic representation of the $v_f$ distribution associated to an obstacle,which is represented by the red line, at fixed time $t$. }
    \label{fig:volumefractions}
\end{figure}

Here we want to give a brief description of how we represent moving obstacles for usage in the ALE-NSB equations.
This task is solved by representing obstacles using triangular surface meshes followed by element-wise calculation of the partial volume covered by the obstacle.
In the first step, all nodes within the obstacle are identified using the ray casting algorithm \cite{Moeller1997, Haines1994}.
Subsequently, all elements are split into three categories and receive a corresponding volume fraction value $v_f$, describing the partial volume covered by the obstacle:
\begin{itemize}
    \item Elements fully covered by the obstacle lie in $\Omega_s$, consequently $v_f=1$.
    \item Elements outside the obstacle lie in $\Omega_f$ and obtain $v_f=0$.
    \item Elements that are split by the element surface correspond to elements in $\Omega_p$, hence
    \begin{equation}
        v_f=\frac{V_{in}}{V_{tot}}
    \end{equation}
    where $V_{in}$ denotes the element volume covered by the obstacle and $V_{tot}$ is the total element volume.
\end{itemize}
This procedure is carried out for every time step
and yields a time-dependent, element-based volume fraction distribution $v_f(t,\vec x)$,
that serves as a basis to provide a suitable permeability distribution, see \autoref{fig:volumefractions}.
In this work we define $\frac{1}{K(t, \vec x)} :=\frac{v_f(t,\vec x)}{\kappa}$ with $\kappa$ being a fixed penalization factor, e.g. $\kappa = 10^{-8}$.
All permeability distributions in this work have been generated using the open-source software \emph{Meshtool}\footnote{\url{https://bitbucket.org/aneic/meshtool/src/master/}}, see \cite{Neic2020} and \cite{Fuchsberger2021} for algorithmic details.
In the case of obstacles that change from open to closed state over time we use a simple scaling function.
For example, assume an obstacle representing a heart valve region will be open at a time instance $t_\mathrm{op}$ and it takes $\mathrm{dur}_\mathrm{V}$ time to switch from open to closed we define
\begin{align}
    \chi(t) := \begin{cases}
                   \frac{t_\mathrm{op}-t}{\mathrm{dur}_\mathrm{V}} & t \in [t_\mathrm{op}-\mathrm{dur}_\mathrm{V},t_\mathrm{op}] \\
                   0 & t \in [t_\mathrm{op}, t_\mathrm{cl} - \mathrm{dur}_\mathrm{V}] \\
                   \frac{t - t_\mathrm{cl}+\mathrm{dur}_\mathrm{V}}{\mathrm{dur}_\mathrm{V}} & t \in [t_\mathrm{cl} - \mathrm{dur}_\mathrm{V},t_\mathrm{cl}] \\
                   1 & \text{else}
               \end{cases},
\end{align}
and modify $\frac{1}{K(t,\vec x)}$ to $\frac{\chi}{K(t,\vec x)}$.

\section{Computation of Residence Times on Moving Domains}
\label{supplrrt}
\begin{figure}
     \centering
     \begin{subfigure}[b]{0.45\textwidth}
         \centering
         \includegraphics[width=\textwidth,keepaspectratio]{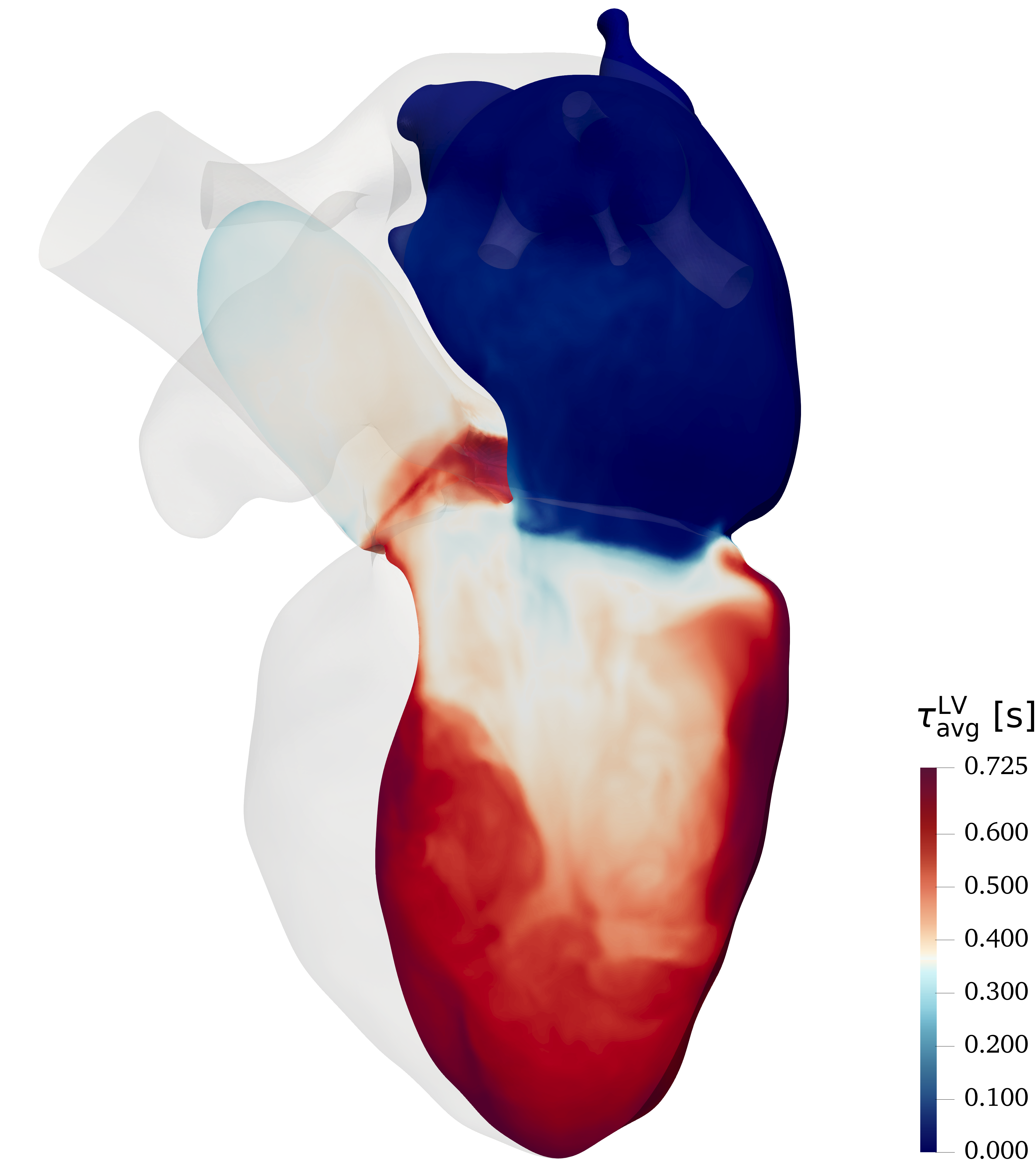}
         \caption{$V(t)$ defined as tag region of left ventricular bloodpool.}
         \label{fig:RTLV}
     \end{subfigure}
     \hfill
     \begin{subfigure}[b]{0.45\textwidth}
         \centering
         \includegraphics[width=\textwidth,keepaspectratio]{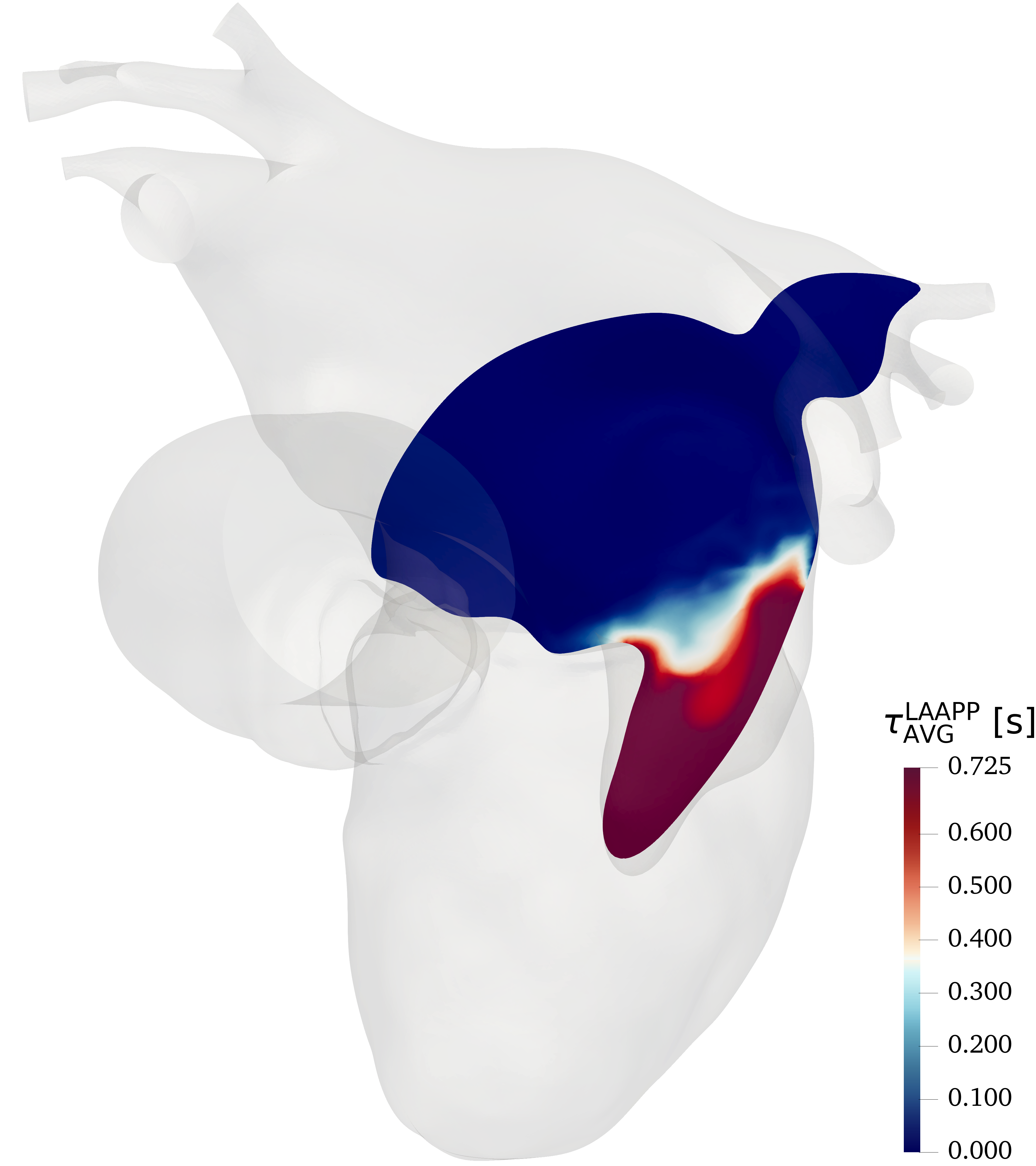}
         \caption{$V(t)$ defined as tag region of left aterial appendage.}
         \label{fig:RTLAAPP}
     \end{subfigure}
        \caption{Time-averaged residence time distributions $\tau_\mathrm{AVG}$ with $V(t)$ defined through different labels in the computational mesh. Time average taken over the final two heartbeats with beatlength equal to \SI{0.725}{\second}.}
        \label{fig:RTLH}
\end{figure}

\begin{figure}
    \centering
    \includegraphics[width=.5\textwidth,keepaspectratio]{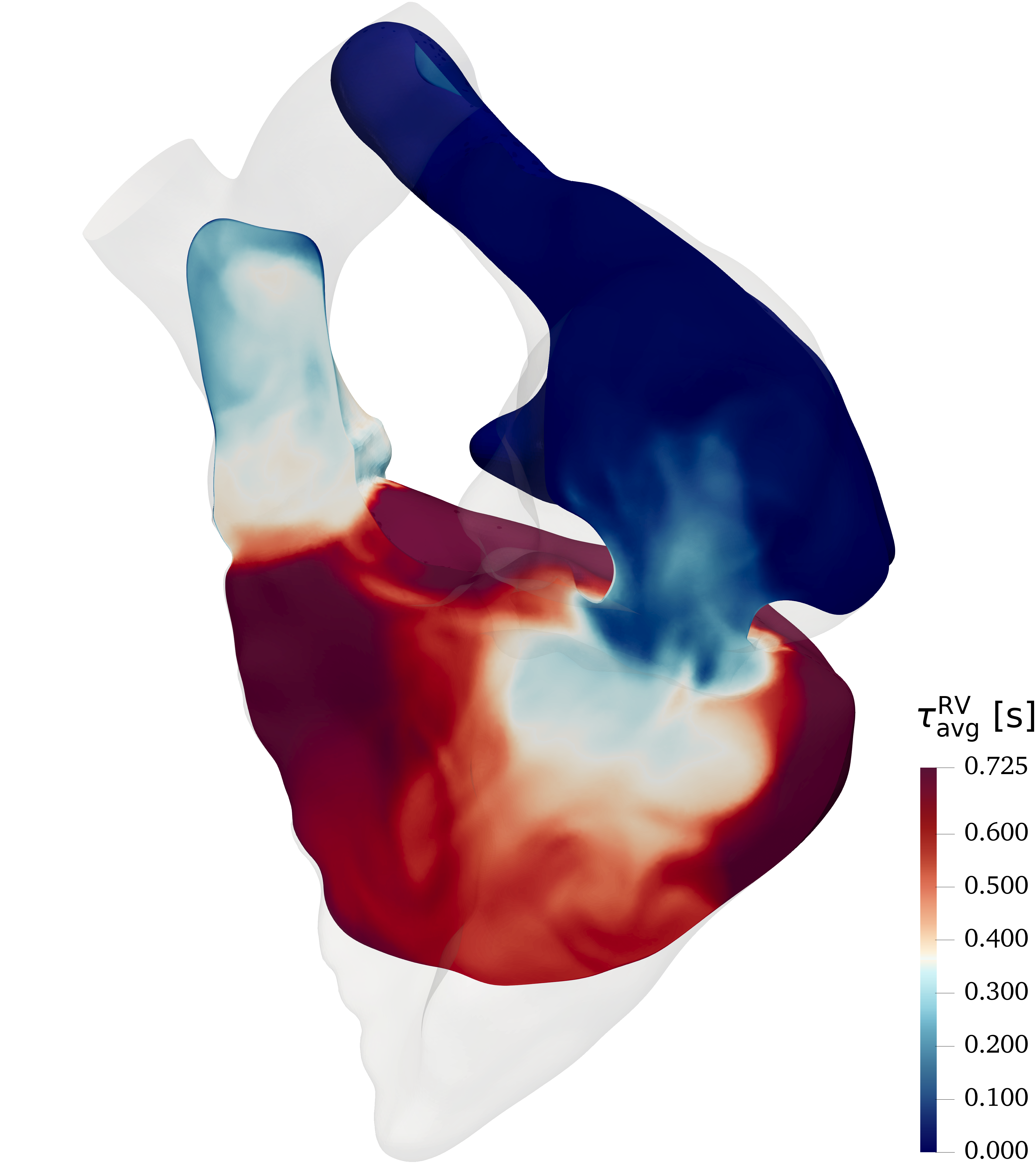}
    \caption{Time-averaged residence time distributions $\tau_\mathrm{AVG}^\mathrm{RV}$ with $V(t)$ defined as right ventricular blood pool in the computational mesh. Time average taken over the final two heartbeats with beatlength equal to \SI{0.725}{\second}.}
    \label{fig:RTRV}
\end{figure}

Here we will give a brief outline of the methods and algorithms used to compute residence time distributions.
The starting point is the following PDE describing the time evolution of a residence time distribution field originating from \cite{EsmailyMoghadam2013,Long2013}. Given a moving fluid domain $\Omega(t) \subset \mathbb R^3$ and a region of interest $V(t) \subset \Omega(t)$ the evolution of the time $\tau(t,\vec x)$ spent in $V$ by an arbitrarily small fluid particle caught at point $\vec x \in \Omega(t)$ at time $t$ can be described as
\begin{align*}
    \frac{\partial }{\partial t} \tau + (\vec u - \vec w) \cdot \nabla \tau - \varepsilon \Delta \tau &= H(t, \vec x) && \text{in } \Omega(t), \\
    \frac{\partial}{\partial \vec n} \tau &= 0 && \text{on } \Gamma_\mathrm{N}(t), \\
    \tau &= g && \text{on } \Gamma_\mathrm{D}(t), \\
    H(t,\vec x) &:= \begin{cases}
                1 & \text{if } (t,\vec x) \in V(t) \\
                0 & \text{else}
             \end{cases},
\end{align*}
with the fluid velocity $\vec u$, the ALE mesh velocity $\vec w$ and a small artificial diffusion parameter $\varepsilon$.
Throughout this work we have used $\varepsilon=\num{1e-12}$.
The fluid velocity $\vec u$ as well as the ALE mesh velocity $\vec w$ are assumed to be given functions, e.g. coming from a pre-computed CFD simulation.
In our applications we set $\Gamma_D(t) = \emptyset$ and $\Gamma_N(t) = \partial \Omega(t)$.
Furthermore, the region of interest $V(t)$ is assumed as a time-evolving tag region assigned to a particular anatomic region, e.g.: ventricular blood pools, and left atrial appendage respectively.
After discretization we have
\begin{align}\label{eq:FCT:1}
    \tensor M_h(t) \dot{\vec \tau} + \tensor K_h(t) \vec \tau(t) = \vec F_h(t)
\end{align}
with
\begin{align*}
    \tensor M_{h,ij}(t) &= \int\limits_{\Omega(t)} \phi_i(t) \phi_j(t) \dd x,\\
    \tensor K_{h,ij}(t)&= \varepsilon\int\limits_{\Omega(t)}\nabla \phi_i(t) \cdot \nabla \phi_j(t)\dd x \\&+ \int\limits_{\Omega(t)}(\vec u(t) - \vec w(t))\cdot \nabla \phi_i(t) \phi_j(t) \dd x, \\
    \vec F_{h,j}(t) &= \int\limits_{\Omega(t)} f(t) \phi_j(t) \dd x,
\end{align*}
with $\{\phi_i(t)\}_{i=0}^n$ being the time-dependent test and trials functions in the ALE domain.
For regular domain movement it is safe to assume that $\tensor M_h(t)$ is invertible for all $t$ and we can rewrite \eqref{eq:FCT:1} as
\begin{align}\label{eq:FCT:2}
    \dot{\vec \tau} + \tensor M_h^{-1}(t)\tensor K_h(t) \vec \tau(t) = \tensor M_h^{-1}(t)\vec F_h(t).
\end{align}
Next, we apply the modified Crank-Nicholson scheme in time as proposed in \cite{Formaggia2004} giving
\begin{equation}\label{eq:FCT:3}
   \left(\tensor M_C + \frac{\Delta t}{2} \tensor K \right)\vec \tau^{n+1} = \left(\tensor M_C - \frac{\Delta t}{2} \tensor K\right)\vec \tau^{n} + \frac{\Delta t}{2} \vec F,
\end{equation}
where we used the following shorthand notation
\begin{align*}
\tensor M_\mathrm{C} &:= \tensor M_h(t^{n+\frac{1}{2}}),\\
\tensor K &:= \tensor K_h(t^{n+\frac{1}{2}}), \\
\vec F &:= \vec F_h(t^{n+\frac{1}{2}}).
\end{align*}
Equation \eqref{eq:FCT:3} is our starting point for applying the FCT scheme similar to \cite{John2008}.
Following the ideas of FEM-FCT methods we define the matrices
\begin{align*}
    \tensor L &:= \tensor K + \tensor D,\\
    \tensor D_{ij} &= -\mathrm{max}\left\{0,\tensor K_{ij}, \tensor K_{ji}\right\} && \text{if } i \neq j,\\
    \tensor D_{ii} &= - \sum_{j=1, j\neq i}^N \tensor D_{ij} && \text{else},\\
    \tensor M_\mathrm{L} &= \mathrm{diag}(\vec m),\\
    \vec m_i &= \sum_{j=1}^N \tensor M_{\mathrm{C},ij}.
\end{align*}
The construction of $\tensor L$ ensures zero row and column sums.
Instead of \eqref{eq:FCT:3} we consider
\begin{equation}\label{eq:FCT:4}
   \left(\tensor M_L + \frac{\Delta t}{2} \tensor L \right)\vec \tau^{n+1} = \left(\tensor M_L - \frac{\Delta t}{2} \tensor L\right)\vec \tau^{n}+ \frac{\Delta t}{2} \vec F,
\end{equation}
which represents a stable low-order scheme whose solution doesn't possess any over or undershoots but suffers from to smeared layers.
To correct this behavior and artificial flux correction vector $\vec f^*(\tau^{n+1}, \tau^n)$ is added to the right hand side of \eqref{eq:FCT:4}.
The definition of $\vec f^*$ follows from an ad-hoc ansatz
\begin{equation*}
    \vec f_{i}^*(\tau^{n+1}, \tau^n)= \sum_{j=1}^n \alpha_{ij}r_{ij},
\end{equation*}
with the \emph{fluxes} $r_{ij}$ defined as
\begin{align}\label{eq:FCT:5}
    r_{ij} &:= \tensor M_{C,ij}(\vec\tau^{n+1}_{i}-\vec\tau^{n+1}_{j})-\tensor M_{C,ij}(\vec\tau^{n}_{i}-\vec\tau^{n}_{j})\\
    \nonumber &-\frac{\Delta t}{2}\tensor D_{ij}(\vec\tau^{n+1}_{i}-\vec\tau^{n+1}_{j})-\frac{\Delta t}{2}\tensor D_{ij}(\vec\tau^{n}_{i}-\vec\tau^{n}_{j}).
\end{align}
and weights $\alpha_{ij} \in [0,1]$.
The representation for $r_{ij}$ follows from first subtracting \eqref{eq:FCT:3} from \eqref{eq:FCT:4} and applying the properties of the matrices $\tensor M_C$ and $\tensor D$.
This formulation represents a nonlinear system.
In \cite{John2008} a linear variant has been proposed which we adapted to the moving-domain case.
For this we use the explicit solution $\tilde{\vec \tau}$ to \eqref{eq:FCT:4}, by means of an explicit Euler scheme approximating the solution $\vec \tau^{n+\frac{1}{2}}$ at time step $t_n + \frac{\Delta t}{2}$, reading
\begin{align*}
    \tilde{\vec \tau} := \vec \tau^n - \frac{\Delta t}{2} \tensor M_L^{-1}\left(\tensor L \vec \tau^n - \vec F\right).
\end{align*}
Inserting $\tilde{\vec \tau}$ into \eqref{eq:FCT:5} and rearranging terms yields
\begin{align*}
    r_{ij} &= \Delta t \left(\tensor M_{C,ij}(\vec \eta^{n+\frac{1}{2}}_{i}-\vec\eta^{n+\frac{1}{2}}_j) - \tensor D_{ij}(\tilde{\vec \tau}_i-\tilde{\vec \tau}_j)\right),
\end{align*}
where $\vec \eta^{n+\frac{1}{2}}_i := (\tensor M_L^{-1}(\vec F - \tensor L \vec \tau^n))_i$.
Additionally, as suggested in \cite{Kuzmin2009}, we employ prelimiting in the form
\begin{equation*}
    r_{ij} = 0 \quad \text{if } r_{ij}(\tilde{\vec \tau}_i - \tilde{\vec \tau}_j) < 0.
\end{equation*}
The computation of the weights $\alpha_{ij}$ follows the same procedure as in \cite{John2008} using Zalesak's algorithm \cite{Zalesak1979}.
We also refer to \cite{Kuzmin2008,Kuzmin2009} for a more detailed overview of the presented method.
Computation of the residence time distribution fields have been included as addon in \emph{CARPentry}.
After computation of the residence time distribution $\tau$ we can calculate the residence time $\mathrm{RT}$ spend in $V(t)$ over a time period $(t_0, t_1)$ \cite{Long2013} as
\begin{align*}
    \mathrm{RT} &:= \frac{1}{(t_1 - t_0) \vert V \vert} \int\limits_{t_0}^{t_1}\int\limits_{\Omega(t)} \tau(t,\vec x) H(t,\vec x)\dx\mathrm{d}t,\\
    \vert V \vert &:= \frac{1}{t_1 - t_0} \int\limits_{t_0}^{t_1}\int\limits_{\Omega(t)}H(t,\vec x)\dx\mathrm{d}t.
\end{align*}
\autoref{fig:RTLH} and \autoref{fig:RTRV} show illustrations of time averaged residence time distributions that were generated for this work as part of the sensitivity analysis.

\section{Pope's Criterion of Turbulence Resolution}\label{sec:pope_crit}
In LES type formulations the resolved velocity field is fundamentally linked to the numerical method used, hence there is no notion of convergence to the solution of a partial differential equation \cite{Pope2004}.
This leads to the problem that mesh convergence often cannot be established by the classical methods \cite{Longest2007,Scuro2018,Jin2017,Hodis2012,Chnafa2014}.
To remedy this problem \cite{Pope2004} proposes the use of a measure of turbulence resolution $M$, see \eqref{eq:TKEfrac}, utilizing the fraction of turbulent kinetic energy resolved by the grid in question.
In order to obtain a point-wise measure, rather than the kinetic energy itself, the kinetic energy density $K(\vec{x},t)$ is considered:
\begin{equation}
    K(\vec{x},t)=\frac{\rho}{2}\vec{u}(\vec{x},t)^2. \label{eq:KE}
\end{equation}
The resulting point-wise measure of turbulence resolution $M$ reads:
\begin{equation}
    M(\vec{x},t)= \frac{K'(\vec{x},t)}{K_{tot}(\vec{x},t)}. \label{eq:TKEfrac}
\end{equation}
Here $K'$ is the turbulent kinetic energy of the residual motions, hence of the motions not resolved by the grid, and $K_{tot}$ stands for the total kinetic energy.
$K_{tot}$ may be written as the sum of the resolved turbulent kinetic energy $K^h$ and the not resolved turbulent kinetic energy $K'$:
\begin{equation}
    K_{tot}(\vec{x},t)=K^h(\vec{x},t)+K'(\vec{x},t)
\end{equation}
The resolved turbulent kinetic energy $K^h$ is calculated from \eqref{eq:KE} using the fluctuating part of the resolved fluid velocity $\bm{u}_f$, which is given by:
\begin{equation}
    \vec{u}_f(\vec{x},t)=\overline{\vec{u}^h}(\vec{x},t)-\vec{u}^h(\vec{x},t),
\end{equation}
where $\overline{\vec{u}^h}$ is an average with respect to time.
When considering a constant inflow $\overline{\vec{u}^h}$ is given by the standard mean over all time steps ($t=1 \dots T$, hence $\overline{\vec{u}^h}$ is not time-dependent) :
\begin{equation}
    \overline{\bm{u}^h}(\bm{x})=\frac{1}{T}\sum_{t=1}^{T}\bm{u}^h(\bm{x},t)
\end{equation}
In the case of a pulsatile behavior however a phase average is considered:
\begin{equation}
    \overline{\vec{u}^h}(\vec{x},t)=\frac{1}{n}\sum_{k=0}^{n-1}\vec{u}^h(\vec{x},t+k \tau),
\end{equation}
where $n$ is the number of cycles or beats and $\tau$ is the period or beat length.
By the use of \eqref{eq:TKEfrac} a criterion for sufficient mesh resolution is given:
\begin{equation}
       M(\bm{x},t) \leq \epsilon_M
\end{equation}
In \cite{Pope2004} a value of $\epsilon_M=0.2$ is proposed, which corresponds to requiring a minimum of 80\% of the total turbulence energy to be resolved.

\section{Input Parameter Variance Effect on Output Features}
\label{supplinterp}
\begin{figure}
    \centering
    \includegraphics[width=\textwidth,keepaspectratio]{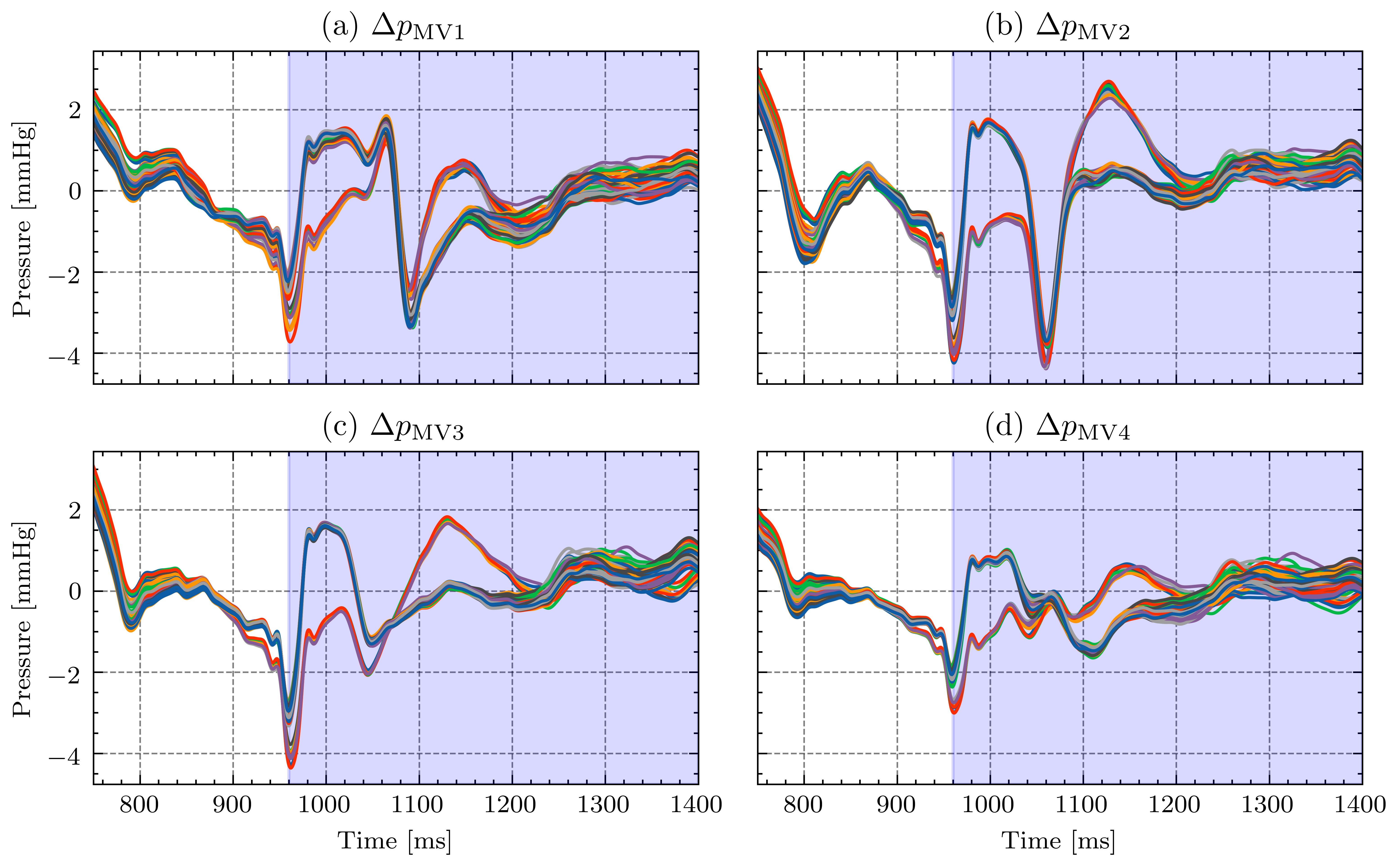}
    \caption{Extracted time signals for pressure differences in the LA for the second heart beat. Diastole is indicated by the shaded blue area in the plots.}
    \label{fig:pressure_overplot}
\end{figure}

This serves as additional interpretation for the results in the main manuscript. \autoref{fig:pressure_overplot} shows the extracted temporal signals of all parameter sets for the pressure differences $\Delta p_{\mathrm{MV}1,2,3,4}$ in the LA.
While there is no strong influence on the output in systole, one can see a clear variation in the outputs in diastole.

\bibliographystyle{abbrvnat}
\bibliography{ms}   %